\documentclass[fleqn,usenatbib]{mnras}

\usepackage{newtxtext,newtxmath}

\usepackage[T1]{fontenc}
\usepackage{ae,aecompl}

\usepackage{graphicx}	
\usepackage{amsmath}	
\usepackage{amssymb}	
\usepackage{verbatim}
\usepackage{soul}
 \usepackage{etoolbox}
 \makeatletter
 \patchcmd\@combinedblfloats{\box\@outputbox}{\unvbox\@outputbox}{}{%
    \errmessage{\noexpand\@combinedblfloats could not be patched}%
 }%
  \makeatother

\title[Recently quenched galaxies in $z~\sim~1$ clusters]{HST/WFC3 grism observations of $z\sim1$ clusters: evidence for evolution in the mass--size relation of quiescent galaxies from poststarburst galaxies}

\author[Matharu et al.]{
\hyperlink{https://orcid.org/0000-0002-7547-3385}{J. Matharu},$^{1,2,3\thanks{E-mail: jmatharu@tamu.edu}}$
\hyperlink{https://orcid.org/0000-0002-9330-9108}{A. Muzzin},$^{4}$
\hyperlink{https://orcid.org/0000-0003-2680-005X}{G. B. Brammer},$^{5}$
\hyperlink{https://orcid.org/0000-0003-1535-2327}{R. F.J. van der Burg},$^{6}$
M. W. Auger,$^{1}$
\newauthor
\hyperlink{https://orcid.org/0000-0002-6528-1937}{P. C. Hewett},$^{1}$
\hyperlink{https://orcid.org/0000-0001-6251-3125}{J. C. C.  Chan},$^{7}$
\hyperlink{https://orcid.org/0000-0003-3921-2177}{R. Demarco},$^{8}$
\hyperlink{https://orcid.org/0000-0002-8282-9888}{P. van Dokkum},$^{9}$
\hyperlink{https://orcid.org/0000-0001-9002-3502}{D. Marchesini},$^{10}$
\newauthor
\hyperlink{https://orcid.org/0000-0002-7524-374X}{E. J. Nelson},$^{11}$
\hyperlink{https://orcid.org/0000-0003-1832-4137}{A. G. Noble}$^{12}$
and 
\hyperlink{https://orcid.org/0000-0002-6572-7089}{G. Wilson}$^{7}$
\\
$^{1}$Institute of Astronomy, University of Cambridge, Madingley Road, Cambridge, CB3 0HA, UK\\
$^{2}$Department of Physics and Astronomy, Texas A\&M University, College Station, TX, 77843-4242, USA\\
$^{3}$Mitchell Institute for Fundamental Physics and Astronomy, Texas A\&M University, College Station, TX, 77845-4242, USA\\
$^{4}$York University, 4700 Keele Street, Toronto, ON, M3J 1P3, Canada\\
$^{5}$Cosmic Dawn Center, Niels Bohr Institute, University of Copenhagen, Juliane Maries Vej 30, DK-2100 Copenhagen \O, Denmark\\
$^{6}$European Southern Observatory, 85748, Garching bei M{\"u}nchen, Germany\\
$^{7}$Department of Physics and Astronomy, University of California Riverside, 900 University Avenue, Riverside, CA 92521, USA\\
$^{8}$Departamento de Astronom\'ia, Facultad de Ciencias F\'isicas y Matem\'aticas,
Universidad de Concepci\'on, Concepci\'on, Chile\\
$^{9}$Astronomy Department, Yale University, 52 Hillhouse Ave, New Haven, CT 06511, USA\\
$^{10}$Department of Physics \& Astronomy, Tufts University, 574 Boston Avenue Suites 304, Medford, MA 02155, USA\\
$^{11}$Harvard-Smithsonian Center for Astrophysics, 60 Garden Street, Cambridge, MA 02138, USA\\
$^{12}$School of Earth and Space Exploration, Arizona State University, Tempe, AZ 85287-1404, USA
\\
}
\date{Accepted XXX. Received YYY; in original form ZZZ}

\pubyear{2019}

\begin{document}
\label{firstpage}
\pagerange{\pageref{firstpage}--\pageref{lastpage}}
\maketitle


\begin{abstract}
Minor mergers have been proposed as the driving mechanism for the size growth of quiescent galaxies with decreasing redshift. The process whereby large star-forming galaxies quench and join the quiescent population at the large size end has also been suggested as an explanation for this size growth. Given the clear association of quenching with clusters, we explore this mechanism by studying the structural properties of 23 spectroscopically identified recently quenched (or ``poststarburst" (PSB)) cluster galaxies at $z\sim1$. Despite clear PSB spectral signatures implying rapid and violent quenching, 87\% of these galaxies have symmetric, undisturbed morphologies in the stellar continuum. Remarkably, they follow a mass--size relation lying midway between the star-forming and quiescent field relations, with sizes $0.1$ dex smaller than $z\sim1$ star-forming galaxies at log$(M_{*}/M_{\odot})=10.5$. This implies a rapid change in the light profile without directly effecting the stellar distribution, suggesting changes in the mass-to-light ratio gradients across the galaxy are responsible. We develop fading toy models to explore how star-forming galaxies move across the mass--size plane as their stellar populations fade to match those of the PSBs. ``Outside-in" fading has the potential to reproduce the contraction in size and increase in bulge-dominance observed between star-forming and PSB cluster galaxies. Since cluster PSBs lie on the large size end of the quiescent mass--size relation, and our previous work shows cluster galaxies are smaller than field galaxies, the sizes of quiescent galaxies must grow both from the quenching of star-forming galaxies and dry minor mergers. 
\end{abstract}




\section{Introduction}

Quiescent galaxies have been observed to grow disproportionately more in size than stellar mass with decreasing redshift (e.g. \citealt{Daddi2005,Trujillo2006,VanDokkum2008,Buitrago2008,VanderWel2008,Damjanov2011,Raichoor2012,Cimatti2012,Mei2012a,Huertas-Company2013,VanderWel2014}). This has been a topic of debate for many studies, with a variety of physical processes suggested for the cause. So far, it has been found that major mergers (equal-mass mergers) lead to a proportionate increase in the stellar mass and size of a galaxy. If this were the mechanism driving galaxy size growth, it would lead to a larger number of high mass galaxies than have been observed \citep{Hopkins2009b,Lopez-Sanjuan2009,Bezanson2009,Naab2009}.

Two processes have emerged that can successfully explain the growth trend of quiescent galaxies with decreasing redshift. The first is the minor mergers (mergers between two galaxies that have a mass ratio of $>$~10:1 or in some cases $>$~3:1) hypothesis of size growth \citep{Bezanson2009,Hopkins2009b,Naab2009,Trujillo2011,Hilz2012,Oser2012}. In this scenario, quiescent galaxies can grow disproportionately more in size than stellar mass, approximately of the form $R_{eff}\propto M_{*}^{2}$ \citep{Naab2009,Bezanson2009,VanDokkum2015}. The second process is the addition of recently quenched galaxies to the quiescent population with decreasing redshift \citep{VanDerWel2009,Carollo2013}. Star-forming galaxies on average have larger sizes than quiescent galaxies at fixed stellar mass and redshift. Furthermore, their sizes at fixed stellar mass increase with decreasing redshift (e.g. \citealt{VanderWel2014}). Hence when star-forming galaxies quench, they join the quiescent population of galaxies at the large size end of the distribution. This process predicts that age gradients as a function of size should exist for quiescent galaxies, whereby larger quiescent galaxies are younger than smaller quiescent galaxies at fixed stellar mass. Several studies have found evidence both for (e.g. \citealt{Fagioli2016,Williams2017}) and against (e.g. \citealt{Whitaker2012a,Keating2014,Yano2016}) this age gradient existing.

Recently, \cite{Matharu2018} used the cluster environment as a laboratory to test whether
minor mergers can explain the majority of the size growth observed in quiescent galaxies residing in the {\it field} environment. Galaxies in high density environments such as clusters have higher peculiar velocities compared to galaxies residing in low density environments. Consequently, satellite galaxies in clusters rarely collide with one another and merge \citep{Merritt1985,Delahaye2017}. This merger suppression (\citealt{Heiderman2009} and references therein) makes growth via minor mergers in clusters a rarity. By measuring the average difference in size between cluster and field galaxies at $z~\mathtt{\sim}~1$ in \cite{Matharu2018}, it was found that cluster galaxies are on average smaller than field galaxies at fixed stellar mass. Other works similar to \cite{Matharu2018} have found a wide range of results. Early-type galaxies at low redshift ($z<0.2$) are often found to exhibit no significant difference in size at fixed stellar mass with environment (e.g. \citealt{Weinmann2009,Maltby2010,Lorenzo2013,Cappellari2013,Huertas-Company2013,Huertas-Company2013a}). However, some studies have found evidence for early-type galaxies being smaller in clusters compared to the field at low redshift (e.g. \citealt{Poggianti2013,Cebrián2014}). At intermediate to high redshifts ($z\geq0.2$), early-type galaxies are more often found to be larger in clusters at fixed stellar mass (e.g. \citealt{Cooper2012,Papovich2012a,Bassett2013,Lani2013,Delaye2014,Chan2018,Andreon2018}). A summary of results from these studies and many others that have been conducted since 2009 are collated in Appendix A of \cite{Matharu2018}. The magnitude of the size difference measured in \cite{Matharu2018} was found to be consistent with the size growth cluster galaxies would have undergone if they had remained in the field rather than accreted on to their clusters at the time most of the mass in their clusters was assembled. This result suggested that minor mergers could indeed be responsible for the dramatic size growth observed in the quiescent field population of galaxies. However, what is now required is evidence for whether or not recently quenched galaxies contribute to this observed size growth.

In this study, we investigate whether the addition of recently quenched cluster galaxies -- also known as ``poststarburst" (PSB) galaxies in the literature -- can lead to an increase in the average size of quiescent cluster galaxies with decreasing redshift. We do this by comparing the position of 23 spectroscopically confirmed recently quenched cluster galaxies on the stellar mass--size relation to the position of the star-forming and quiescent cluster members. Our size measurements for the cluster sample come from {\it Hubble Space Telescope} (HST), {\it Wide Field Camera 3} (WFC3) F140W imaging of ten clusters at $z~\mathtt{\sim}~1$ drawn from the GCLASS survey. This sample represents the largest sample of spectroscopically confirmed cluster galaxies at $z~\mathtt{\sim}~1$ to-date.

This paper is organised as follows. In Section~\ref{sample} we describe our data and the sample of recently quenched cluster galaxies. Section~\ref{direct_images} discusses the F140W direct images of the recently quenched galaxies and what clues this provides for the quenching mechanism(s) responsible. The position of the recently quenched cluster galaxies on the stellar mass--size relation is presented and studied in Section~\ref{PSBS_mass_size}. A more thorough analysis of their morphology with respect to other cluster members is then studied in Section~\ref{morphology_sec}. In Section~\ref{model}, we investigate if the distinct properties of the recently quenched galaxies in GCLASS can aid in explaining their evolutionary path on the mass--size plane. We discuss the implications of our findings in Section~\ref{discussion} and summarise in Section~\ref{summary}.

All magnitudes quoted are in the AB system and we assume a $\Lambda$CDM cosmology with $\Omega_{m}=0.307$, $\Omega_{\Lambda}=0.693$ and $H_{0}=67.7$~kms$^{-1}$~Mpc$^{-1}$ \citep{Planck2015}.

\section{Data}
\label{sample}
\subsection{Cluster sample}
\label{cluster_sample}
The cluster sample from which the PSBs are identified consists of 10 massive ($M_{200}>10^{14}M_{\odot}$) clusters in the redshift range $0.86<z<1.34$. These clusters are a subsample of the clusters discovered in the 42 square degree {\it Spitzer} Adaptation of the Red-sequence Cluster Survey (SpARCS, see \citealt{Muzzin2009}; \citealt{Wilson2009a} and \citealt{Demarco2010}). This subsample of clusters was selected for spectroscopic follow-up as part of the Gemini Cluster Astrophysics Spectroscopic Survey (GCLASS, see \citealt{Muzzin2012} and \citealt{VanderBurg2013}). As part of this survey, extensive optical spectroscopy was obtained using the Gemini Multi-Object Spectrographs (GMOS) on both Gemini-South and -North. In total, 1282 galaxies obtained a spectroscopic redshift, with 457 being identified as cluster members. There is also 11-band photometry available for the ten clusters ({\it ugrizJK$_{s}$}, 3.6$\mu$m, 4.5$\mu$m, 5.8$\mu$m, 8.0$\mu$m), details of which can be found in Appendix A of \mbox{\cite{VanderBurg2013}}. The study presented in this paper is based on newly acquired photometry with HST in the F140W (wide {\it JH} wavelength range of \mbox{$12003<~\lambda~/$~{\AA}~$<15843$}) filter.

\subsection{HST data}
\label{HST_data}

\begin{figure*}

	\centering\includegraphics[width=\textwidth]{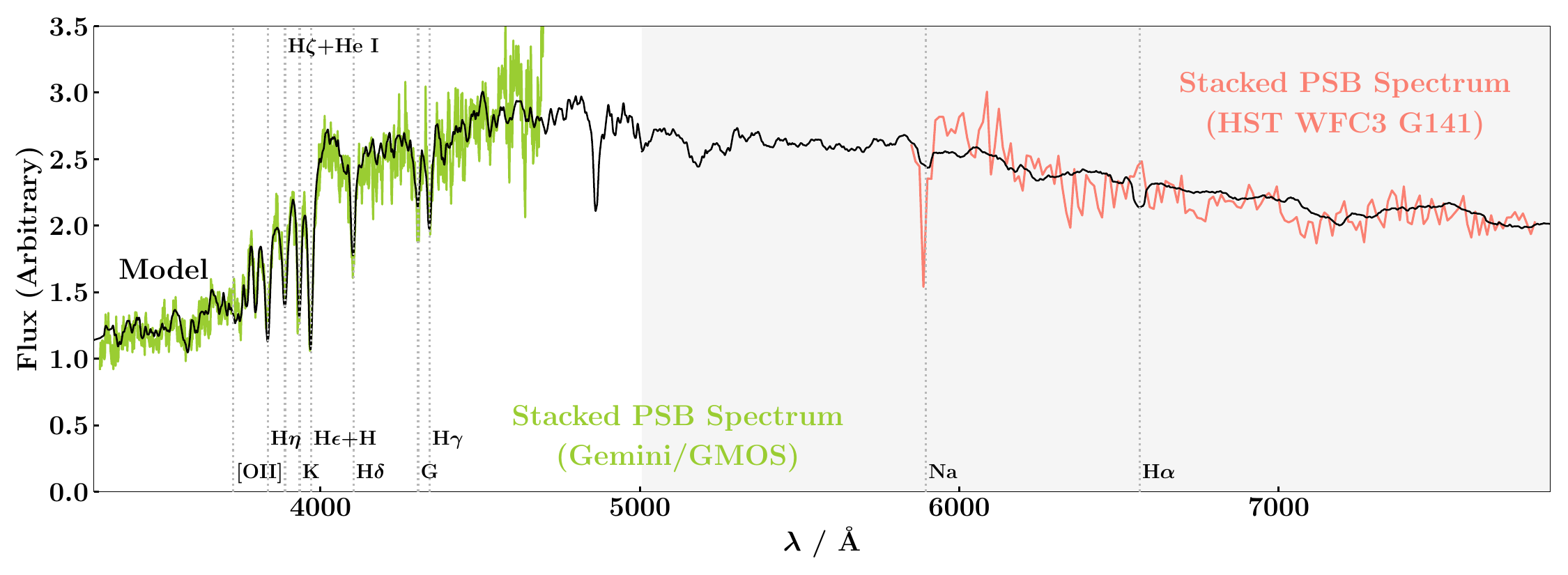}
    \caption{Normalised mean stacked spectra for the GCLASS PSBs from Gemini/GMOS in the optical (shown in green) and from the HST WFC3 G141 grism in the infrared (shown in red). Stacked spectrum in the optical is calculated from 28 PSBs. Stacked spectrum in the infrared is calculated from 13 PSBs (13 of the 23 in the HST fields-of-view had good quality grism spectra). Overlaid black spectrum shows the best-fit \protect\cite{Bruzual2003} spectrum \protect\citep{Muzzin2014a}. This model has been smoothed with a uniform filter such that it has the approximate resolution of GMOS ($R=450$) up to $\sim$5000{\AA}, after which it has the approximate resolution of the G141 grism ($R=130$). The part of the plot in which the model has been smoothed to the resolution of the G141 grism has been shaded with a light grey background. Prominent emission/absorption lines have been labeled with dotted grey vertical lines. [OII] is {\it not} detected, but its position is labeled for reference.}
    \label{fig:psb_spectrum}
\end{figure*}

A HST WFC3 F140W imaging and G141 grism follow-up of the 10 clusters from GCLASS was conducted to obtain spatially resolved H$\alpha$ maps of the star-forming cluster galaxies (GO-13845; PI Muzzin). The wavelength coverage of the G141 grism spans \mbox{$10750<~\lambda~/$~{\AA}~$<17000$}. The  H$\alpha$ emission line can therefore be detected for galaxies between \mbox{$0.7<z<1.5$}. A spatially resolved spectrum for every object in the field of view can be obtained with a resolving power of \mbox{$R~\mathtt{\sim}~130$}.

These data were used to carry out a detailed comparison of the stellar mass--size relation in the clusters compared to a field population of galaxies from 3D-HST at the same redshift (see \citealt{Matharu2018}). 392 of the 457 spectroscopically confirmed cluster members in GCLASS are in the HST fields-of-view, of which 344 met the sample selection criteria for the stellar mass--size relation study in \cite{Matharu2018}. An additional 177 cluster members were identified using the grism spectra. 5 galaxies with low-confidence spectra from GCLASS were also confirmed as cluster members using grism spectra. 
This amounted to a final sample of 526 cluster galaxies. For more details on the HST data, cluster member identification process using G141 spectra and the sample selection criteria for the stellar mass--size relation study, we refer the reader to \cite{Matharu2018}.

The stellar mass--size relation study presented in \cite{Matharu2018} allowed for the structural analysis of the GCLASS cluster galaxies for the first time: and with it, the PSBs -- the results of which we will present in this paper.

\subsection{PSB sample}
\label{PSB_sample}

The PSBs form part of the sample of cluster galaxies that were spectroscopically confirmed as part of the GCLASS survey (see Section~\ref{cluster_sample}). As part of a study into the effects of environment and stellar mass on galaxy properties, the PSBs were identified as recently quenched star-forming galaxies based on features in their GMOS spectra. These were identified from 9 of the 10 clusters, since the rest-frame wavelength coverage of the spectra for the 10th cluster were limited to $\lambda_{rest}< 4050${\AA} due to its high redshift. The spectra were taken with the R150 grating, giving a spectral resolution of $R=450$. In the rest-frame, this corresponds to a spectral resolution of $\sim8${\AA}, and is sufficient for identifying absorption lines. The wavelength coverage of the spectra are \mbox{$6300~<~\lambda/${\AA}~$<~9200$}, corresponding to \mbox{$3000~\lesssim~\lambda/${\AA}$~\lesssim~4500$} in the rest-frame. For full details on the GMOS spectra, we refer the reader to \cite{Muzzin2012}.

\subsubsection{PSB classification}
\label{psb_classification}
The criteria for the poststarburst classification was chosen to select galaxies similar to K+A galaxies which are identified from rest-frame optical spectra in the literature \citep{Dressler1999,Poggianti1999, Balogh1999,Poggianti2009,Yan2009}. Despite there being no strict definition for these galaxies, they are usually defined by their spectral properties, as having an absence of emission lines, but strong Balmer line absorption (EW(H$\delta$)$\geqslant5${\AA}). The H$\delta$ absorption feature that is required in the K+A classification has weak equivalent widths of 2-7{\AA}. The signal-to-noise ratio of most galaxies in GCLASS at $z\sim1$ make it difficult to detect reliably. This line is also further weakened by the contamination of many sky lines. Consequently for GCLASS, $D(4000)$ was used since it is correlated with EW(H$\delta$) \citep{Balogh1999}. $D(4000)$ is defined as the strength of the 4000{\AA} break in spectra. For the GMOS PSB spectra, it is measured by taking the ratio of the flux in the red continuum at 4000-4100{\AA} to that in the blue continuum at 3850-3950{\AA} \citep{Balogh1999,Muzzin2012}. The GCLASS PSBs were therefore defined as those galaxies with an absence of [OII] emission and  $1.0<~D(4000)<1.45$ in their GMOS spectra. A stacked spectrum of the cluster galaxies that met this criteria revealed a spectrum very similar to that of a K+A galaxy, suggesting that the average galaxy selected by this criteria is a K+A galaxy. For more details on the PSB classification process, we refer the reader to \cite{Muzzin2012}. 

\subsubsection{PSB completeness}

When studying galaxy populations in spectroscopic surveys, there are certain aspects that are more important to consider in comparison to imaging surveys. Unlike imaging surveys, spectroscopic surveys cannot necessarily yield data for all galaxies of interest in the field-of-view. Masks need to be made prior to observation, requiring slit placement to be predetermined by some criteria. This therefore means spectroscopic surveys are almost always incomplete.

The completeness of a spectroscopic sample quantifies how many objects of a particular classification have spectroscopy versus how many objects of the same classification exist in the field-of-view within the target of interest (such as a galaxy cluster for example). For samples like the PSB sample, which are classified solely by their spectroscopic properties, it is challenging to determine the true size of the parent PSB sample. In such scenarios, it becomes more important to ensure the sample is not biased in any way which makes it unrepresentative of the parent population. If the sample can be deemed representative, and therefore a likely reflection of the physical properties the parent population would possess, we can make meaningful conclusions regarding the physical properties of the entire PSB cluster population.

In the context of studying the mass--size relation properties of the PSBs, our PSB sample is incomplete, but should be representative of the underlying PSB population. Since the most reliable way to identify PSBs is via spectroscopy, it is difficult to know the parent population of possible PSBs that could have been targeted for spectroscopy in GCLASS. Nevertheless, even if the true underlying population of PSBs spans a wide range in colour, we still expect our PSB sample to be representative of the parent population. This is because the slit placement for the GCLASS GMOS spectroscopy used a very broad colour cut (see Figure 2 and Section 3.1 of \cite{Muzzin2012} for more details). Indeed, the PSB classification (Section~\ref{psb_classification}) led to an almost equal number of UVJ- star-forming and quiescent galaxies within the PSB sample (see Figure~\ref{fig:mass_size_PSBs}). This is an indication that the PSBs were sampled in a representative way, since we naturally expect the colours of PSBs to be intermediate between those of UVJ- star-forming and quiescent cluster galaxies.

Consequently, any missing PSBs from our PSB sample would not significantly alter our measured PSB mass--size relation (Figure~\ref{fig:mass_size_PSBs}) or the conclusions we make about morphology in Section~\ref{morphology_sec}.

\subsubsection{PSB stacked spectra}

The stacked spectrum from GMOS, along with the stacked spectrum obtained from the G141 grism for the PSBs is shown in Figure~\ref{fig:psb_spectrum}. The grism spectra were taken with the WFC3 G141 grism on board the HST. These spectra have a resolution of $R=130$ and a wavelength coverage of \mbox{$10750~<~\lambda~/$~{\AA}~$<~17000$}. This corresponds to a rest-frame coverage of \mbox{$5300~\lesssim~\lambda/$~{\AA}$~\lesssim~8500$}. Raw grism spectra are two-dimensional, providing a spatially-resolved spectrum of each object in the field-of-view (see Figure 1 of \cite{Matharu2018} for example). One-dimensional spectra can be obtained from these and stacked. 13 of the 23 PSBs presented in this study had grism spectra with good enough quality (no contamination from neighbouring objects and more than half the grism spectrum available) such that they could be stacked. 28 PSBs were used for the stacked spectrum in the optical, of which 5 are not in the HST fields-of-view and therefore not part of the study presented in this paper. It is clear from Figure~\ref{fig:psb_spectrum} that the optical spectra (shown in green) fit the model (shown in black) well. Despite the much lower signal-to-noise ratio of the grism spectra (shown in red), they remain consistent with the optical template. There is also evidence for very weak H$\alpha$ emission, since H$\alpha$ emission is clearly seen in the grism spectrum of one of the PSBs. Nevertheless, the absence of H$\alpha$ emission for most of the PSBs confirms that these galaxies have strongly suppressed star formation.

\subsubsection{Distinction between high and low redshift PSBs}

Since the classification of the GCLASS PSBs was based on a spectroscopic criteria that was designed to apply to $z\sim0$ galaxies, a clarification is required as to what galaxies this selection criteria leads to at $z\sim1$, the redshift of GCLASS. In general, galaxies in the nearby universe have much lower star formation rates (SFRs) than those at higher redshifts. To explain a PSB spectrum at low redshifts, the galaxy in question must have experienced a starburst before it quenched. i.e, it must have had high levels of star formation, which are uncommon for $z=0$ star-forming galaxies. A galaxy undergoing a starburst usually lies systematically above the star-forming main sequence at fixed redshift, with SFRs~$\gtrsim4$ (e.g. \citealt{Noeske2007,Rodighiero2011, Elbaz2017}). Towards higher redshifts, the intercept of the star-forming main sequence (SFR--stellar mass relation for star-forming galaxies) increases \citep{Noeske2007,Whitaker2012b}. Therefore, a galaxy that is considered a starburst galaxy at $z=0$ will be considered a {\it normal} star-forming galaxy at higher redshifts. The GCLASS PSBs reside at $z\sim1$, where SFRs were $\sim6$ times higher\footnote{At a fixed stellar mass of Log$(M_{*}/M_{\odot})\sim10$.} than at $z\sim0$ \citep{Whitaker2012b,Whitaker2014}. Employing the term ``poststarburst" is somewhat misleading in connection with this sample of galaxies because these are normal star-forming galaxies at $z\sim1$ that have had their regular levels of star formation suddenly terminated, rather than true PSB galaxies akin to those found locally. There is no requirement to invoke a starburst as the explanation for their spectral properties. At this redshift, they can therefore be considered as ``post star-forming" galaxies. Despite the term ``poststarburst'' being misleading for this sample of galaxies, we choose to continue using it to maintain consistency with previous analyses on this sample \citep{Muzzin2012,Muzzin2014a} and with the K+A notation used throughout the literature.

\section{HST imaging of the PSBs}
\label{direct_images}
\begin{figure}
	\centering\includegraphics[width=\columnwidth]{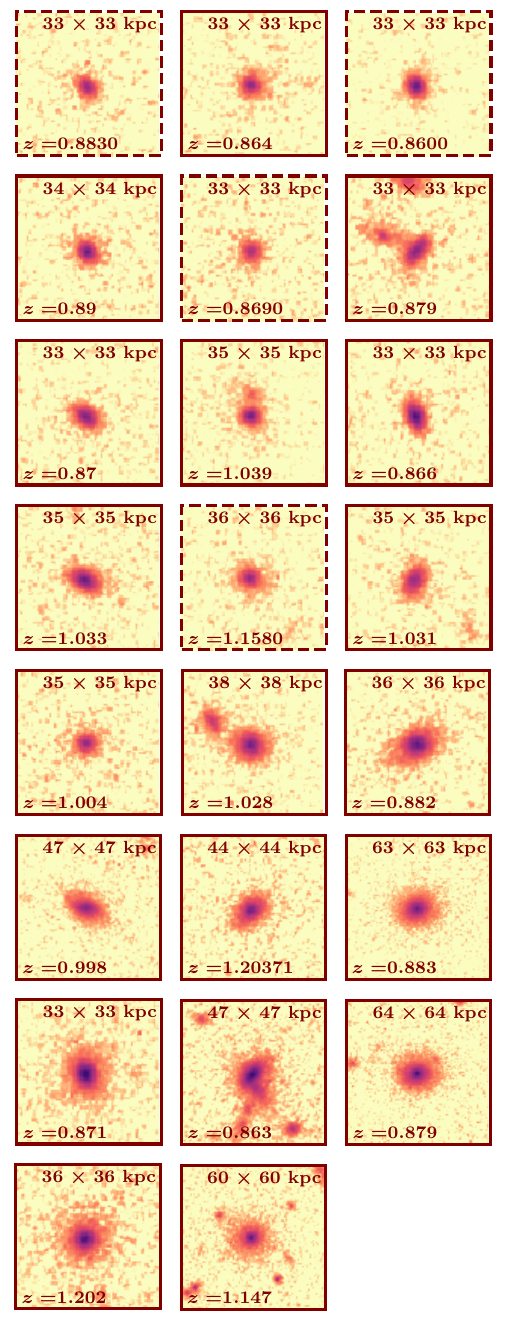}
    \caption{HST WFC3 F140W cutouts of the GCLASS PSBs in order of increasing stellar mass. PSBs with reliable (low-confidence) spectroscopic redshifts are shown with solid (dashed) line borders. The dimensions of each cutout in kiloparsecs and the spectroscopic redshift of each PSB is shown in the top right- and bottom left-hand corners of each cutout respectively. The colourmap is logarithmic.}
    \label{fig:psb_collage}
\end{figure}

In Figure~\ref{fig:psb_collage} we show the F140W direct images of the PSBs in square cutouts. The dimensions of each cutout in kiloparsecs is shown in the top right-hand corner of each cutout. The spectroscopic redshift of each galaxy is shown in the bottom left-hand corner of each cutout. The images are ordered by increasing stellar mass. Cutouts with solid line (dashed line) borders are those PSBs with reliable (low-confidence) spectroscopic redshifts.

The majority of the PSBs show no signs of disturbances or interactions and are symmetric in shape. This is somewhat puzzling, since the abrupt end in these galaxies' star formation suggests that a severe violent process (e.g. a merger or a high-speed close-proximity flyby, both of which would likely leave tidal signatures or some sort of asymmetry) would likely be responsible. The fact that most of the PSBs are symmetric and undisturbed effectively rules out violent quenching mechanisms that are capable of altering the stellar distribution of galaxies. These include harassment, mergers and tidal-stripping. To further investigate this, we examine the residuals of the GALFIT fits for the PSBs (see Figure~\ref{fig:GALFIT_PSBs_fig} in Appendix~\ref{GALFIT_psbs}). The majority of the residuals are absent of features that are suggestive of harassment, mergers or tidal-stripping.

These violent quenching mechanisms were already ruled out in more detailed previous work done on the PSBs. A detailed study using the large sample of spectroscopically confirmed cluster galaxies in GCLASS unveiled that the PSBs reside in a distinct location of clustercentric velocity versus position phase space \citep{Muzzin2014a}. Using several zoom-in dark-matter-only simulations, the coherent ``ring" traced by the PSBs in this phase space could be reproduced by quenching galaxies rapidly (within 0.1 - 0.5 Gyr) after they made their first passage of $R\sim0.5R_{200}$. In \cite{Muzzin2014a} it was also found that the GMOS spectra of the PSBs could only be fit well with high resolution \cite{Bruzual2003} stellar population models if a rapid quenching timescale of $0.4^{+0.3}_{-0.4}$~Gyr is assumed. The rapidity of the quenching, suggested {\it both} by this phase space analysis and the Spectral Energy Distribution (SED) fitting of the PSBs, ruled out quenching mechanisms that operate on long timescales such as mergers, harassment and tidal-stripping. The short quenching timescale is consistent with gas stripping processes such as ram-pressure stripping \citep{Gunn1972} and/or strangulation \citep{Larson1980}.

Some of the PSBs however do show evidence of processes involving the possible stripping of stars or mergers. The most striking example is the second PSB on the penultimate row, which seems to have a tail of stripped stars as well as asymmetry in the overall galaxy. The last PSB on the second row shows evidence of a bridge between itself and a close companion. A close companion is also seen near the second PSB on the fifth row, which may be tidally interacting with the PSB. Whilst some of the PSBs have clear neighbouring galaxies, none have identifiable tidal features. Additionally, it is reasonable to expect close companions in projection when observing high density environments such as clusters. Furthermore, the PSBs quenched $\sim0.5$~Gyr prior to observation \citep{Muzzin2014a}. If a merger, harassment or tidal interaction was responsible for triggering the PSB phase, $0.5$~Gyr is not enough time for the stars to settle into a symmetric, undisturbed disc. Therefore, it seems unlikely that mergers or tidal stripping are the dominant processes responsible for the emergence of the GCLASS PSBs.

\section{PSBs on the cluster stellar mass--size relation}
\label{PSBS_mass_size}

\begin{figure*}
	\includegraphics[width=\textwidth]{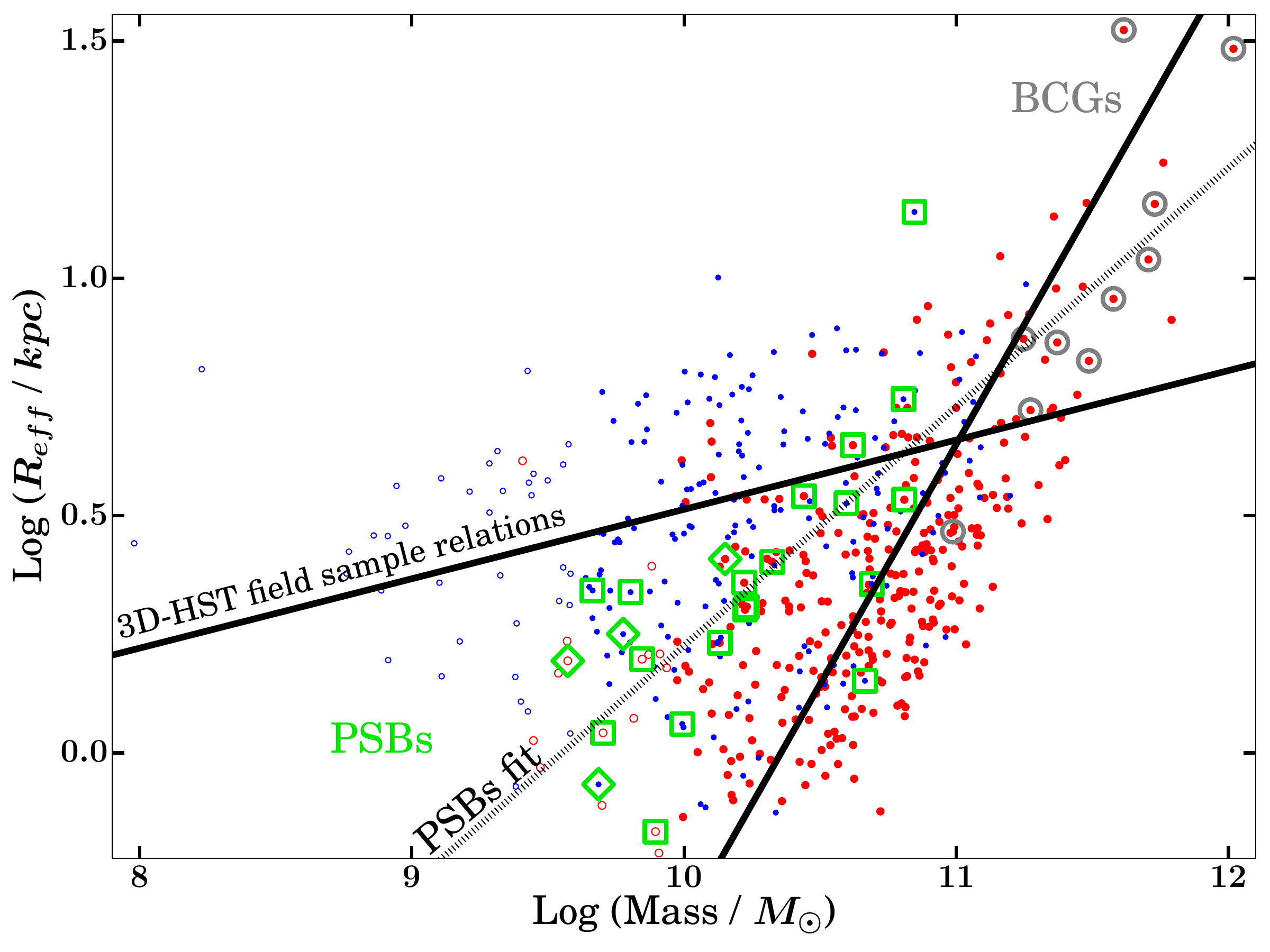}
    \caption{Stellar mass--size relation for the GCLASS clusters. Blue (red) filled circles show star-forming (quiescent) galaxies within the mass completeness limits. $R_{eff}$ is the half-light radius in kpc. Brightest Cluster Galaxies (BCGs) are circled in grey. Poststarburst (PSBs) galaxies with reliable spectroscopic redshifts are marked with green squares. PSBs with low-confidence spectroscopic redshifts are marked with green diamonds. Solid black lines show the field relations at $z~\mathtt{\sim}~1$ calculated using results from \protect\cite{VanderWel2012,VanderWel2014}. Open blue (red) circles show measurements for star-forming (quiescent) galaxies beyond the mass completeness limits.}
    \label{fig:mass_size_PSBs}
\end{figure*}

Figure~\ref{fig:mass_size_PSBs} shows the cluster stellar mass--size relation for all 10 of the GCLASS clusters with the PSBs highlighted in green. Sizes are the half-light radii measurements along the semi-major axis from single-component S\'ersic profile fits with GALFIT \citep{Peng2002b, Peng2010a}. Details of the size determination process and its reliability can be found in \cite{Matharu2018}. The PSBs that are highlighted with green squares have reliable spectroscopic redshifts, and were therefore included in the cluster mass--size relation study for GCLASS in \cite{Matharu2018}. The PSBs highlighted by green diamonds have low-confidence spectroscopic redshifts which resulted in them being excluded from the analysis in \cite{Matharu2018}. Nevertheless, we include them on this plot for reference, since they have properties consistent with the PSBs that have higher quality spectra. Open blue (red) circles indicate star-forming (quiescent) cluster galaxies below the mass completeness limits of GCLASS. The mass completeness limits of GCLASS are log$(M_{*}/M_{\odot})=9.96$ and log$(M_{*}/M_{\odot})=9.60$ for quiescent and star-forming cluster galaxies respectively. For details on how these were calculated, we refer the reader to \cite{Matharu2018}. Galaxies were classified as star-forming or quiescent based on their rest-frame $UVJ$ colours (see \citealt{Matharu2018}). The solid black lines show the field relations at $z~\mathtt{\sim}~1$ calculated using stellar mass and size measurements from 3D-HST (see \cite{Matharu2018} for a detailed explanation on how these were calculated). 

It can be clearly seen that the PSBs do not follow the same stellar mass--size relation as the star-forming or quiescent galaxies. They seem to lie in a distinct region, halfway between the star-forming and quiescent mass--size relations. We would like to emphasise that the PSBs were selected based solely on their spectroscopic properties and not on any visual property such as morphology. The PSB sample of galaxies in GCLASS were defined in \cite{Muzzin2012,Muzzin2014a}, well before the HST imaging (Section~\ref{HST_data}) was obtained. Now, after measuring their stellar masses and sizes, we have found they also lie on a distinct mass--size relation, indicative of galaxies transitioning from star-forming to quiescent-like structural properties.
We therefore choose to fit a relation to the PSBs, shown as the black dotted line. This is a least-squares fit to the PSBs with reliable spectroscopic redshifts. 

It is clear that the PSBs do not follow the mean mass--size relation of star-forming galaxies for their redshift. This is remarkable, because they have only {\it recently} quenched. These galaxies must have been on the star-forming mass--size relation a few hundred million years prior to $z\sim1$. Since the PSBs have only recently quenched, it is not expected that they would have had enough time to alter their structural properties significantly from those of typical star-forming galaxies. Previous work on the PSBs has shown that they likely began quenching $\sim0.5$~Gyr prior to observation (see Section~\ref{direct_images} and \citealt{Muzzin2014a} for more details). Most of the PSBs have symmetric, undisturbed morphologies in the stellar continuum (see Section~\ref{direct_images}). They also have transitional stellar mass--size properties between those reminiscent of star-forming and quiescent galaxies. Together, these results seem to suggest that the quenching mechanism responsible affects solely the gas content of galaxies, and not the stars. However, it is somehow capable of altering the light profile of a galaxy in a very specific way.

\section{Morphology of the PSBs}
\label{morphology_sec}

\begin{figure*}
	\includegraphics[width=\textwidth]{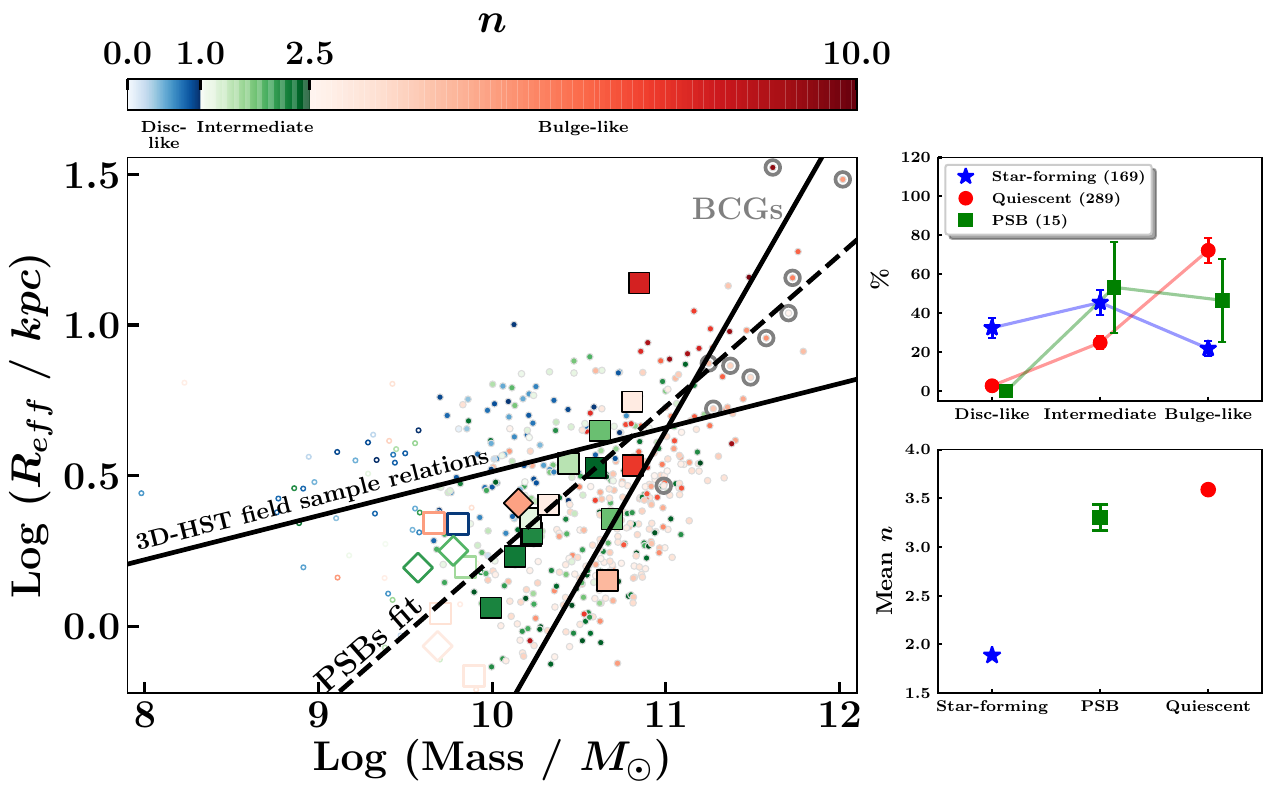}
    \caption{Main panel: GCLASS stellar mass--size relation with points colour-coded by S\'ersic index. Colour-coding corresponds to the three bins used for morphological indication in \protect\cite{Matharu2018}. Gradient colourmaps are used within each morphological class to highlight to what degree a galaxy is either disc-like, intermediate-type or bulge-like. PSBs with reliable spectroscopic redshifts are shown as large squares. PSBs with low-confidence spectroscopic redshifts are shown as large diamonds. All other cluster galaxies are shown as small circular points in the background. Open markers show measurements below mass completeness limits. Lines are the same as in Figure~\ref{fig:mass_size_PSBs} and the BCGs are circled in grey. Top right panel: The percentage of each morphological type within the quiescent, star-forming and PSB (offset for clarity) classifications. These calculations only take into account galaxies within the mass completeness limits, of which the sample sizes are stated in the legend within brackets. Error bars are Poisson errors. Bottom right panel: The average S\'ersic index of the star-forming, PSB and quiescent populations for galaxies within the mass completeness limits. Error bars show the standard error in the mean.}
    \label{fig:morphology}
\end{figure*}

Figure~\ref{fig:morphology} shows the cluster stellar mass--size relation in the largest panel, but now with the points colour-coded by their S\'ersic index, $n$. Following previous work done on the morphological variation of cluster galaxies across the mass--size plane \citep{Matharu2018}, we divide S\'ersic index measurements into three bins. Disc-like, intermediate-type and bulge-like galaxies are defined as those galaxies with $n\leqslant1$, $1<n<2.5$ and $n\geqslant2.5$ respectively. These are shown in blue, green and red respectively. Within each of these three morphological classifications, we show to what degree a galaxy is either disc-like, intermediate-type or bulge-like by using gradient colourmaps. In general, lower S\'ersic indices are a proxy for more disc-like morphologies and vice versa. PSBs with reliable (low-confidence) spectroscopic redshifts are shown as large squares (diamonds). The smaller circular points in the background show the rest of the cluster galaxies. BCGs are circled in grey. The lines are the same as in Figure~\ref{fig:mass_size_PSBs}. The top right panel shows the percentage of each morphological type within the star-forming, quiescent and PSB (offset for clarity) populations within the mass completeness limits. The quiescent mass completeness limit is used for the PSBs and the error bars show Poisson errors. The bottom right panel shows the mean S\'ersic index for each of the three populations within the mass completeness limits. The errors bars show the standard error in the mean.

Since S\'ersic index has been found to be well-correlated with quiescence (e.g. \citealt{Franx2008,Bell2012}), it is not surprising that we find the quiescent (star-forming) population with the highest percentage of bulge-like (disc-like) galaxies. This can be seen both in the main panel -- where most of the blue (red) circular points lie on the star-forming (quiescent) mass--size relation -- and in the top right panel. Also evident from the top right panel is that the PSBs have the highest percentage ($\sim50\%$) of intermediate-type galaxies out of the three populations. They also have a percentage of bulge-like galaxies that falls almost exactly between the percentages of bulge-like galaxies found for the star-forming and quiescent populations. Interestingly, they have the lowest percentage of disc-like galaxies among the three populations. Due to poor sample statistics (stated in the legend of this panel), the uncertainties on the PSB morphology fractions are large \mbox{($\sim \pm20\%$)}. We therefore appreciate that these results need to be verified with better statistics in future work. Since the PSB population is dominated by intermediate-types and bulge-like galaxies, it is therefore not surprising that the PSBs have a mean S\'ersic index that is closer to the mean S\'ersic index of the quiescent population, yet still intermediary between the mean S\'ersic indices of the star-forming and quiescent populations (bottom right-hand panel).

Since the PSBs have only recently quenched, we would have expected them to have a similar percentage of disc-like galaxies to the star-forming population and therefore a mean S\'ersic index much closer to that of the star-forming population. Instead, these results suggest that the PSBs are galaxies which morphologically transformed both recently and rapidly.

The PSBs do not exhibit disturbed morphologies in the stellar continuum (see Section~\ref{direct_images}), but do have S\'ersic indices more reminiscent of bulge-like morphologies. Since the PSBs are recently quenched galaxies, it is surprising how quickly their morphology has become bulge-dominated. One hypothesis is that before these galaxies quenched, they had both a disc and bulge component, but the disc is likely to have been brighter than the bulge. The rapid quenching could have caused a rapid fading of the disc, such that the bulge became relatively brighter than the disc. Quenching processes that are capable of causing a rapid fading of galactic discs without affecting the stellar distribution are processes which solely affect the gas content of galaxies. Candidate processes for this are ram-pressure stripping \citep{Gunn1972} and strangulation/starvation \citep{Larson1980}. The gas -- which is required for star formation -- is more loosely bound in galactic discs. Therefore, these processes remove gas more efficiently from the discs of galaxies rather than their bulges \citep{Abadi1999}.

In the next section, we investigate whether a simple disc-fading model of galaxy quenching can explain the formation of most of the GCLASS PSB population.

\section{Are PSBs faded discs?}
\label{model}

As mentioned at the end of the previous section, the quenching processes that are likely to be responsible for the emergence of the PSBs can cause the fading of galactic discs. To build a better understanding of how disc-fading affects the structural properties of typical star-forming galaxies at $z\sim1$, we will be presenting a variety of disc-fading toy models in this Section. For each model, we will track how typical star-forming galaxies move across the mass--size plane as their discs fade, and how their morphology (as dictated by the S\'ersic index) changes.

The star-forming, PSB and quiescent cluster galaxies in GCLASS have very well defined stellar populations as dictated by their spectral properties (e.g. see Section~\ref{PSB_sample} for a discussion on the spectral properties of the PSBs). Therefore, there are tight constraints on the allowed amount of disc-fading in each model, such that the final galaxy exhibits stellar populations reminiscent of the average PSB.

The aim of this Section is therefore twofold: we will be exploring whether disc-fading can simultaneously create a population of galaxies with similar stellar populations to the PSBs {\it and} similar structural properties to the PSBs, as dictated by their distinct mass--size relation.

\subsection{Disc-fading toy model (bulge and disc)}
\label{model2}
\begin{table*}
	\centering
	\caption{Summary of parameters for model galaxies in the disc-fading toy model presented in Section~\ref{model2}. $n$ is S\'ersic index, $b/a$ is axis ratio and P.A. is position angle measured in degrees counter-clockwise.}
	\label{tab:brightness_table}
	\begin{tabular}{llccccccc}
    	\hline
        Galaxy & Component & $D(4000)$ & Combined & Contribution to & Magnitude & $n$ & $b/a$ & P.A.\\
         & & & $D(4000)$ & total brightness (\%) & & & & \\
        \hline\hline
        Starting & Bulge & 1.7 & 1.2 & 17 & 23.75 & 4 & 0.76 & 90\\
         & Disc & 1.09 & & 83 & 22.00 & 1 & 0.62 & 90\\
         \hline
         Faded & Bulge & 1.7 & 1.4 & 33 & 23.75 & 4 & 0.76 & 90\\
          & Disc & 1.25 &  & 67 & 22.75 & 1 & 0.62 & 90\\
          \hline
	\end{tabular}
\end{table*}
The first model we explore is one in which galaxies are modelled as having both a bulge and disc component. In typical star-forming galaxies, galactic bulges tend to host the oldest stars in the galaxy. Therefore, in our model, we ensure that the bulge models have a brightness that reflects older stellar populations than the disc models.

Since the stellar populations of the GCLASS cluster galaxies are well defined (Section~\ref{model}), there are strict constraints on the brightness of the bulge and disc components in this model. Similarly, there are tight constraints on how much the disc can fade such that the resulting galaxy has stellar populations reminiscent of the average PSB.

The first step in applying these constraints requires an understanding of how the brightness of a stellar population varies with age at fixed stellar mass. To do this, we determine the F140W stellar mass-to-light ratios as they would be observed at $z~\mathtt{\sim}~1$ for a set of stellar population models with solar metallicity using the \cite{Bruzual2003} libraries. A \cite{Chabrier2003} initial mass function is used. We then see how these mass-to-light ratios vary with $D(4000)$ (see Figure~\ref{fig:ml_ratios_plot} in Appendix~\ref{ml_ratios}), which is a good indicator for age (see Section~\ref{PSB_sample}).

\subsubsection{Relative brightnesses of the bulge and disc}
\label{relative}

In the GCLASS spectroscopic sample (see Section~\ref{cluster_sample}), disc-dominated ($n\leqslant1$) star-forming cluster galaxies have an average $D(4000)=1.2$, the PSBs have an average $D(4000)=1.4$ and bulge-dominated ($n\geqslant2.5$) quiescent cluster galaxies have an average $D(4000)=1.7$. In our disc-fading model, we require our starting galaxies to resemble star-forming disc-dominated galaxies and our faded galaxies to resemble the PSBs. Therefore, we use the average $D(4000)$ values for these two populations in GCLASS to determine the $D(4000)$ values of the disc components for the starting and disc-faded galaxies in our model. For simplicity, we create model galaxies {\it only} for the case where the stellar mass of the bulge equals the stellar mass of the disc. For the bulge components, we use $D(4000)=1.7$, the average $D(4000)$ of bulge-dominated quiescent cluster galaxies in GCLASS. We then find the corresponding F140W mass-to-light ratio for $D(4000)=1.7$ from the relation between $D(4000)$ and the F140W stellar mass-to-light ratios in our stellar population models (see Figure~\ref{fig:ml_ratios_plot} in Appendix~\ref{ml_ratios}). The $D(4000)$ value for the disc component that ensures the {\it overall} $D(4000)=1.2$ for the starting galaxy is then found, which is $1.09$. Similarly, the $D(4000)$ value for the disc component that ensures the overall $D(4000)=1.4$ for the faded galaxy is found, which is $1.25$. These are found by calculating the weighted mean (based on the relative brightness of the disc and bulge from their F140W mass-to-light ratios) of the $D(4000)$ values for the bulge and disc components in each case. These parameters are summarised in Table~\ref{tab:brightness_table}. For the starting galaxy models, the disc is 5 times brighter than the bulge. For the faded galaxy models, the disc is 2 times brighter than the bulge. Therefore, the disc is faded by a factor of $\sim2$ in our disc-fading model.

\subsubsection{S\'ersic profiles of model galaxies}
\label{profiles}

\begin{figure*}
	\includegraphics[width=\textwidth]{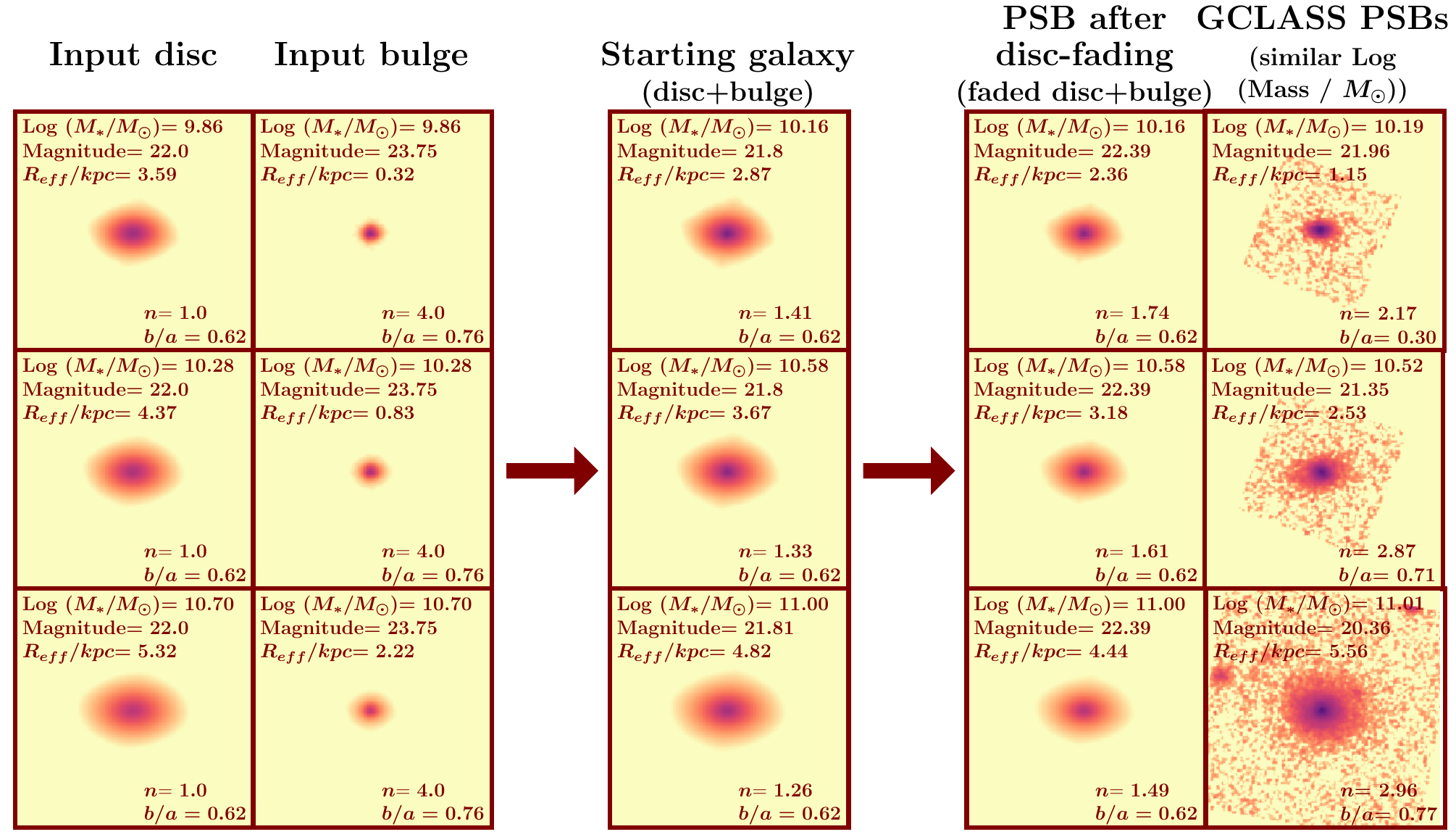}
    \caption{Examples of models created with GALFIT for the disc-fading toy model (Section~\ref{model2}). Model galaxies for three stellar masses (Log$(M_{*}/$M$_{\odot})=10.16, 10.58$ and $11.00$) are shown (third and fourth column). First and second columns show the disc and bulge models, respectively. The discs follow the $z~\mathtt{\sim}~1.1$ star-forming disc mass--size relation in the field. The bulges follow the $z~\mathtt{\sim}~1$ quiescent field mass--size relation (see Section~\ref{profiles}). All disc models have Magnitude$=22.0$, S\'ersic index, $n=1.0$ and axis ratio, $b/a=0.62$. All bulge models have a magnitude that is 5 times fainter than the disc models (see Section~\ref{relative}). All bulge models are set to have $n=4.0$ and $b/a=0.76$. Bulge and disc models have the same stellar masses. The disc and bulge models are then combined to create starting galaxies shown in the third column. The disc model is then faded by a factor of 2 and combined with the input bulge model to create faded disc galaxies, shown in the penultimate column. Final column shows PSBs from GCLASS with similar stellar masses to the galaxy models. Parameters listed in all cutouts show the measured values from GALFIT. For the models, sizes are converted into kpc assuming the models are at $z=1$. The dimensions of all cutouts are the same, as well as the position angle (measured counter-clockwise) of all models and the PSBs, set to 90 degrees. The colourmap is logarithmic.}
    \label{fig:fading_models}
\end{figure*}

Model S\'ersic profiles for the discs and bulges are created using GALFIT \citep{Peng2002b,Peng2010a}. These have the same properties as the GCLASS F140W images (e.g. resolution and magnitude zeropoint). Since the PSBs began quenching $\sim0.4$~Gyr ago \citep{Muzzin2014a}, the discs of the starting galaxies in the disc-fading toy model must follow the mass--size relation for disc-dominated star-forming galaxies at this epoch, which corresponds to $z\sim1.1$. We calculate this mass--size relation using a large sample of disc-dominated star-forming galaxies in the field at $z\sim1.1$, taken from 3D-HST (\citealt{VanderWel2012}). These are calculated with the same method\footnote{Except that in this case there is one sample. We run the fitting method 1000 times on this sample to capture the range of possible intercept and gradient values. The average value for the intercept and gradient are used for the final relation.} used to calculate the $z\sim1$ field relations (Section~\ref{PSBS_mass_size}). We assume that the PSBs will eventually evolve into quiescent galaxies that are similar to the quiescent bulge-dominated population in GCLASS. Quiescent bulge-dominated galaxies in 3D-HST at $z\sim1$ have a mass--size relation that is consistent with the $z\sim1$ field mass--size relation for the entire quiescent population (see for example Figure 9 in \citealt{Matharu2018}). Therefore, our bulge models have stellar masses and sizes that follow the $z\sim1$ quiescent field mass--size relation. We create disc and bulge models spanning the stellar mass range of the GCLASS PSBs ($9.32<$~Log$(M_{*}/M_{\odot})<11$).

We assign axis ratios to the disc and bulge models that are typical of star-forming and quiescent field galaxies at $z\sim1$ (see Figure 4 of \citealt{Hill2019}). Disc and bulge models are assigned S\'ersic indices typical of disc-dominated star-forming and bulge-dominated quiescent galaxies, respectively (see Figure 9 of \citealt{Matharu2018}). All disc models have a S\'ersic index $n=1$ and an axis ratio $b/a=0.62$\footnote{This is the axis ratio of $UVJ$-selected star-forming galaxies at z~$\sim~1$ with $M_{*}=10^{11}M_{\odot}$ in 3D-HST+CANDELS, interpolated from values calculated for $0.5<z<1.0$ and $1.0<z<1.5$ shown in Figure 4 of \citealt{Hill2019}.}. All bulge models have $n=4$ and $b/a=0.76$\footnote{This is the axis ratio of $UVJ$-selected quiescent galaxies at z~$\sim~1$ with $M_{*}=10^{11}M_{\odot}$ in 3D-HST+CANDELS, interpolated from values calculated for $0.5<z<1.0$ and $1.0<z<1.5$ shown in Figure 4 of \citealt{Hill2019}.}. 

The magnitudes of all the discs are set to F140W magnitude=22, which is the average magnitude of the star-forming disc-dominated cluster galaxies in GCLASS. The bulges are 5 times fainter than the discs for the starting galaxies (F140W magnitude=23.75) and the discs are 2 times brighter (F140W magnitude=22.75) than the bulges for the faded galaxies. A summary of all the important parameters for the disc and bulge models is shown in Table~\ref{tab:brightness_table}.

Model galaxies are then created by adding the corresponding disc and bulge model S\'ersic profiles. Examples of galaxy models for three stellar masses within the mass completeness limits of GCLASS (Section~\ref{PSBS_mass_size}) are shown in Figure~\ref{fig:fading_models}. Alongside the disc-faded galaxy models, we show select PSBs that have a similar stellar mass to the disc-faded model galaxies.

\subsubsection{The stellar mass--size relation of faded discs}
\label{faded_ms}
\begin{figure*}
	\includegraphics[width=\textwidth]{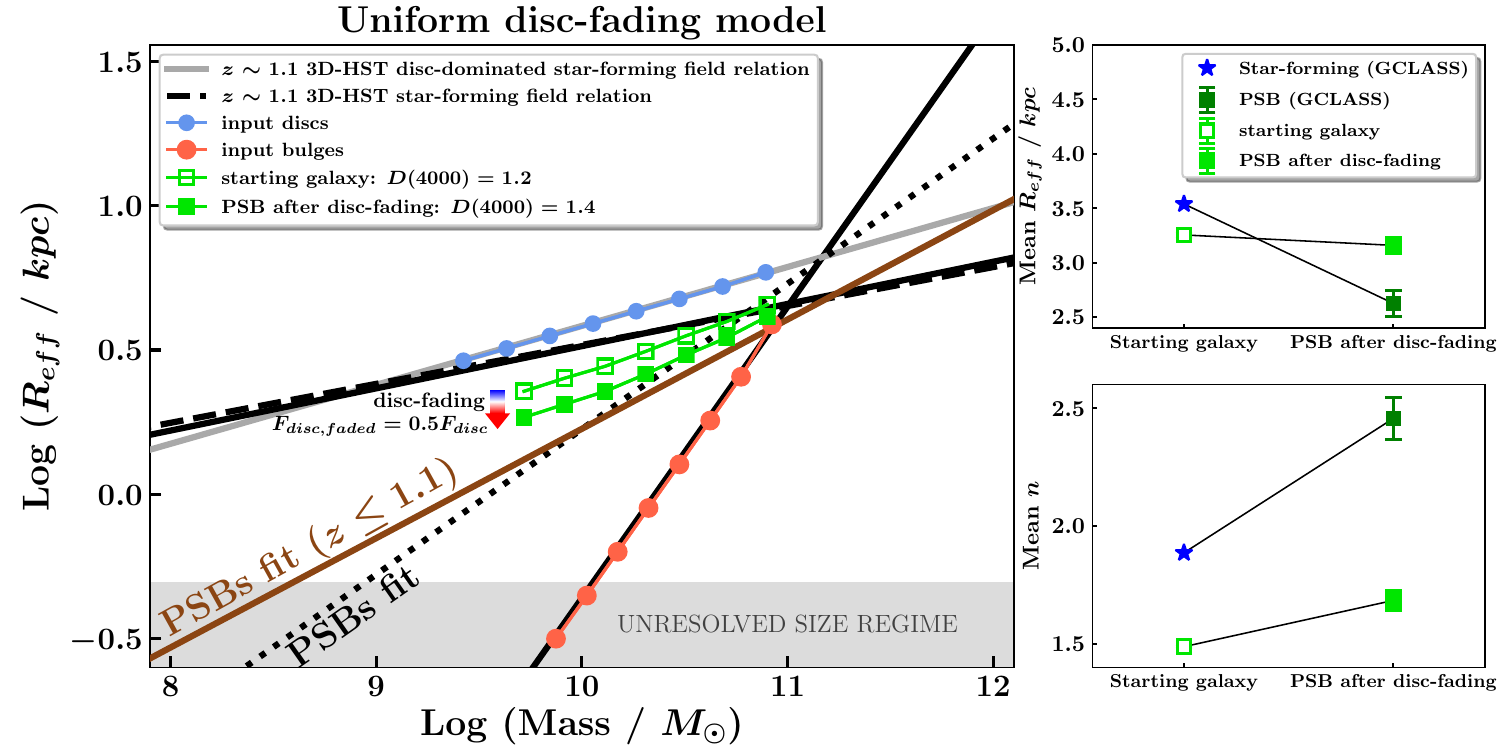}
    \caption{Disc-fading toy model. Main panel: median size measurements for 8 stellar mass bins of the disc (small blue circles) and bulge (large red circles) input S\'ersic models that were created to follow the $z\sim1.1$ disc-dominated star-forming field mass--size relation and the $z\sim1$ quiescent field mass--size relation, respectively. For reference, we show the $z\sim1.1$ star-forming field mass--size relation as the dashed black line. The $z\sim1$ field mass--size relations are the solid black lines (the same solid black lines as in Figures~\ref{fig:mass_size_PSBs} and \ref{fig:morphology}) and the dotted black line is the same PSB fit shown in Figure~\ref{fig:mass_size_PSBs}. Bulges are five times fainter than the discs at fixed stellar mass (see text for reasoning). The starting galaxy is composed of the disc+bulge input models, median size measurements of which in seven stellar mass bins are shown as open green squares. Corresponding median size measurements of the disc-faded galaxies are shown as filled green squares. Since $M_{*, bulge} = M_{*, disc}$, the stellar masses of the starting and disc-faded galaxies are $2M_{*,bulge}$ (or $2M_{*,disc}$). The disc is faded by a factor of 2 (see text for reasoning). The region in which size measurements fall below the resolution limit of the GCLASS F140W images is shaded in grey. Top and bottom right panels: average half-light radius and S\'ersic index values for the starting and disc-faded galaxy models along with the values for the GCLASS star-forming and PSB cluster galaxies (see text for more details). Error bars on GCLASS measurements are standard errors in the mean. Error bars on the model results are standard deviations from monte carlo sampling (see text for details).}
    \label{fig:fading_toy_model}
\end{figure*}

The S\'ersic models have their sizes and S\'ersic indices determined using the same two-GALFIT-run approach described in \cite{Matharu2018}. The same point-spread function (PSF) is used for all the models. This is just one of the pre-selected stars (see \citealt{Matharu2018}) in the WFC3 F140W images of one of the GCLASS clusters. A noise map is not used since our models are noiseless (apart from the noise introduced by convolving the models with the PSF such that they match the resolution limit of WFC3) and no sky estimation is carried out since there is no sky background in our models.

The resulting median size measurements for multiple stellar mass bins (\mbox{$9.32<$~Log$(M_{*}/M_{\odot})<11$} for discs, \mbox{$9.80<$~Log$(M_{*}/M_{\odot})<11$} for bulges and \mbox{$9.62<$~Log$(M_{*}/M_{\odot})<11$} for galaxies)\footnote{Bulge model measurements are only plotted within the quiescent mass completeness limit, since lower stellar mass results fall well within the unresolved size regime (see Figure~\ref{fig:fading_toy_model}). Since galaxy models have stellar masses $2M_{*, bulge}$ (or $2M_{*, disc}$), the minimum stellar mass for which there are galaxy models is Log$(M_{*}/M_{\odot})=9.62$. We only plot model results up to the intersection point of the $z~\sim~1$ field mass--size relations.} are shown as large points on the mass--size relation in the largest panel of Figure~\ref{fig:fading_toy_model}. The disc and bulge input model measurements are shown as small blue and large red circles respectively. The resolution limit of the F140W images is equivalent to 0.5 kiloparsecs at $z=1$. Therefore, the sizes measured for the bulge models that fall below this threshold are unresolved and therefore unreliable. This regime of unresolved sizes is shown as the shaded grey region. The median size measurements of the starting star-forming galaxies are shown as open green squares. The median size measurements of the faded disc galaxies are shown as filled green squares. The dashed black line shows the $z\sim1.1$ star-forming field relation\footnote{This is calculated in exactly the same way as the $z\sim1.1$ mass--size relation for disc-dominated star-forming galaxies in the field (Section~\ref{profiles} and the solid grey line in Figure~\ref{fig:fading_toy_model}), except all star-forming field galaxies at $z\sim1.1$ are used.}, which is expected to approximately follow the mass--size relation of the starting galaxies. The solid grey line is the $z\sim1.1$ mass--size relation for disc-dominated star-forming galaxies in the field (see Section~\ref{profiles}). The solid black lines show the $z\sim1$ field relations (which are the same solid black lines in Figures~\ref{fig:mass_size_PSBs} and \ref{fig:morphology}) for reference.

It is evident that disc-fading leads to a reduction in the overall size of a galaxy at fixed stellar mass. However, given the tight constraints on the allowed amount of fading from the measured $D(4000)$ values, the magnitude of this reduction is not enough to explain the stellar mass--size relation of the PSBs. Since our starting galaxies are assumed to be star-forming at $z\sim1.1$, it is impossible to have PSBs with $z>1.1$ in our models\footnote{If we neglect PSBs with $z>1.1$ when calculating the relevant fading factors in any of our toy models, our conclusions in Section~\ref{model} remain unchanged.}. Therefore, we recalculate the PSB mass--size relation using {\it only} those PSBs at $z\leqslant1.1$ with reliable spectroscopy redshifts. In this disc-fading model and all subsequent disc-fading models (Sections~\ref{outsidein_model} and \ref{bulge_disc_outsidein_model}), we compare the relative difference between this relation (shown as the solid brown line in the largest panel of Figures~\ref{fig:fading_toy_model}, \ref{fig:outsidein_fading_toy_model} and \ref{fig:bulge_disc_outsidein_fading_toy_model}) and the $z\sim1.1$ star-forming field mass--size relation to the relative difference between our starting galaxies and disc-faded galaxies on the mass--size plane. The reason the starting galaxies do not lie on the $z\sim1.1$ star-forming field mass--size relation is because their discs are not bright enough. If the discs were brighter than they currently are, the bulge components would become even more subdominant in brightness, leading to an increase in the overall half-light radius of the starting galaxy model. Consequently, the position of the starting galaxy models on the mass--size plane would be in better agreement with the $z\sim1.1$ star-forming field mass--size relation. There are two ways in which this can be improved. The first is to increase the disc stellar mass (and therefore by definition the size due to the mass--size relation) for each starting galaxy. When testing this in the modelling, we found that a more disc-dominated starting galaxy led to a smaller size reduction as a result of disc-fading. The second option is to reduce the stellar age of the disc, thereby increasing its brightness. This increases the difference in stellar age between the starting and faded galaxies, thereby increasing the size reduction due to disc-fading. However, reducing the average age of the stellar populations in the disc violates our $D(4000)$ constraints from the observations. Therefore, the results shown in Figure~\ref{fig:fading_toy_model} show the optimal performance of our disc-fading model under our tight constraints. The magnitude of the size drop, shown by the length of the blue-to-red coloured arrow, is much smaller than the size difference between the $z\sim1.1$ star-forming field mass--size relation and the $z\leqslant1.1$ PSB mass--size relation. Therefore, uniform disc-fading is incapable of bringing the faded disc galaxies inline with the $z\leqslant1.1$ PSB mass--size relation. Quantitatively, at the median stellar mass of the PSBs residing at $z\leqslant1.1$ that fall within the quiescent mass completeness limit (Log$(M_{*}/$M$_{\odot}) = 10.39$), the half-light radius shrinks by 16\% as a result of disc-fading.

In the top right-hand panel of Figure~\ref{fig:fading_toy_model}, we show the average sizes of the starting and faded galaxies within the GCLASS mass completeness limits, as well as the average sizes of the star-forming and PSB galaxies in GCLASS. For GCLASS, the averages for the star-forming galaxies are calculated by taking all the star-forming galaxies within the star-forming mass completeness limit, and then calculating the mean size. The average for the PSB galaxies is calculated by taking all the PSB galaxies with reliable spectroscopic redshifts at $z\leqslant1.1$ that are beyond the quiescent mass completeness limit, and then calculating the mean size. The starting galaxy models have exactly the same stellar masses as the GCLASS star-forming cluster galaxies beyond the star-forming mass completeness limit. The disc-faded models have exactly the same stellar masses as the PSBs with reliable spectroscopic redshifts at $z\leqslant1.1$ that are beyond the quiescent mass completeness limit. The errors on the model means are standard deviations in the mean size measurements from monte carlo sampling. We run the disc-fading model 10 times for model galaxies. Each time, we add a 10\% random uncertainty to the size, drawn from a normal distribution. We then calculate the mean size in each run and the standard deviation in the mean sizes calculated from the 10 runs are the error bars shown. For GCLASS measurements, the errors are standard errors in the mean.

It is clear from the top right-hand panel of Figure~\ref{fig:fading_toy_model} that disc-fading does lead to a drop in the average size of a galaxy, however, this drop is much smaller than observed between the average sizes of the star-forming and PSB galaxies in GCLASS.

\subsubsection{The average S\'ersic index of faded discs}
\label{sersic}

In the bottom right-hand panel of Figure~\ref{fig:fading_toy_model}, we compare the average S\'ersic indices of our starting and disc-faded galaxies in the toy model to the average S\'ersic indices of the GCLASS star-forming cluster members and PSBs. For GCLASS, the averages for the star-forming galaxies are calculated by taking all the star-forming galaxies within the star-forming mass completeness limit, and then calculating the mean S\'ersic index. The average for the PSB galaxies is calculated by taking all the PSB galaxies with reliable spectroscopic redshifts at $z\leqslant1.1$ that are beyond the quiescent mass completeness limit, and then calculating the mean S\'ersic index. For the starting galaxy models, the mean S\'ersic indices are for model galaxies that have exactly the same stellar masses as the star-forming cluster galaxies in GCLASS beyond the star-forming mass completeness limit. For the disc-faded model galaxies, the mean S\'ersic indices are for model galaxies that have exactly the same stellar masses as the PSBs with reliable spectroscopic redshifts at $z\leqslant1.1$ that are beyond the quiescent mass completeness limit. The errors are calculated from
the same monte carlo sampling described in Section~\ref{faded_ms}. For GCLASS measurements, the errors are standard errors in the mean.

It is expected that in the event of disc-fading, the bulge will become more prominent. This is also reflected in Table~\ref{tab:brightness_table}, where the bulge contributes to the total brightness of the galaxy by a larger amount in the disc-faded case. Since bulge-dominated galaxies have on average higher S\'ersic indices than disc-dominated galaxies, it is not surprising that we see a rise in the average S\'ersic index as a result of disc-fading in the toy model. However, the difference between the average S\'ersic index of the faded disc galaxies and the starting galaxies is smaller than the difference seen in the observations. The larger difference in the average S\'ersic index of the PSBs relative to the star-forming galaxies in GCLASS suggests that disc fading alone is not responsible for the structural properties of the PSBs. Another process is required that is capable of enhancing the brightness of the bulge.

\subsection{Disc-fading toy model (outside-in fading of a disc-only model)}
\label{outsidein_model}

\subsubsection{Properties of the starting and faded galaxies}
\label{outside_inprop}
In Section~\ref{model2}, we presented a disc-fading model whereby the galaxy was modelled as having a bulge and disc component. The disc was uniformly faded such that the resulting disc plus bulge model resembled the stellar populations of the GCLASS PSBs. While this uniform fading of the disc did make galaxies smaller overall, the magnitude of this size reduction did not lead to galaxies that resembled the structural properties of the PSBs.

Observations of ram-pressure stripping at low redshift have shown that environmental quenching happens in an ``outside-in" fashion (e.g. \citealt{Koopmann2004a,Boselli2006,Abramson2011a}). Star formation in environmentally-quenched galaxies is truncated from the outside-in, such that over time, it is confined further and further towards the centre of the galaxy. This leads to truncated H$\alpha$ discs in cluster galaxies. Recently, more detailed observations of ram-pressure stripping have led to the quantification of quenching timescales and ages as a function of galactocentric radius (\citealt{Fritz2017,Bellhouse2017,Gullieuszik2017,Jaffe2018,Vulcani2018,Fossati2018,Cramer2019}). Since the outskirts of the disc is quenched first, there is a gradient in the stellar populations of a galaxy undergoing ram-pressure stripping, such that the outskirts hold older stellar populations than the central region. Therefore, the uniform fading of the disc in the disc-fading model presented in Section~\ref{model2} does not capture how disc-fading as a result of ram-pressure stripping operates.

To improve our model of disc-fading such that it is representative of how ram-pressure stripping operates, we present an alternative disc-fading model in this Section, where the disc is faded differentially, such that the outskirts are faded more than the central regions. As was the case in the previous model, the amount of fading is constrained by the $D(4000)$ measurements of the star-forming and PSB galaxies in GCLASS.

For this disc-fading model, we do not model the galaxy with bulge and disc components. We model the starting galaxies with a single disc component, since a typical star-forming galaxy is disc-dominated (see Figure~\ref{fig:morphology} for example). The brightness of the starting galaxy models is dictated by the stellar populations of a typical disc-dominated ($n\leqslant1$) star-forming cluster galaxy in GCLASS. At fixed stellar mass, the average disc-dominated star-forming cluster galaxy in GCLASS has $D(4000)=1.2$. After disc-fading, our galaxy models must have a brightness similar to that of the average PSB in GCLASS. The average PSB has $D(4000)=1.4$. As was done in the previous disc-fading model, we refer to our stellar populations models to find out the difference in brightness of a stellar population with $D(4000)=1.2$ and $D(4000)=1.4$ at fixed stellar mass in the F140W filter. As a reminder, we do this by determining the F140W stellar mass-to-light ratios as they would be observed at $z\sim1$ for a set of stellar population models with solar metallicity using the \cite{Bruzual2003} libraries (see Figure~\ref{fig:ml_ratios_plot} in Appendix~\ref{ml_ratios}). A \cite{Chabrier2003} initial mass function is used. We then see how these mass-to-light ratios vary with $D(4000)$, which is a good indicator for age (see Section~\ref{PSB_sample}).

\subsubsection{Outside-in fading}
\label{outsidein_fading}

The difference in brightness between a stellar population with $D(4000)=1.2$ and $D(4000)=1.4$ tells us the approximate difference in brightness between a typical disc-dominated star-forming galaxy and PSB in GCLASS. In our outside-in fading model, this difference in brightness will correspond to how much the central regions of our starting galaxy models will be faded by. It is particularly important that the central regions of the faded galaxies reflect the stellar populations of a typical PSB. This is because the spectra from which the PSB properties are measured depend mostly upon light from this region of the galaxy. The central regions of a galaxy are the brightest, and therefore contribute most to the measured spectrum. A stellar population of solar metallicity with $D(4000)=1.4$ (typical PSB) is a factor 1.5 fainter than a stellar population of solar metallicity with $D(4000)=1.2$ (typical disc-dominated star-forming cluster galaxy). Therefore, the central regions of the starting galaxy models will be faded by a factor of 1.5.

For a guide on the required levels of fading for the outskirts of the galaxy models, we use the results presented in \cite{Fossati2018} for NGC 4330. NGC 4330 is a galaxy in the Virgo cluster that is undergoing ram-pressure stripping. Using multiwavelength imaging, \cite{Fossati2018} were able to confirm ``outside-in" quenching as well as quantify quenching times as a function of galactocentric radius for this galaxy. They find that ``outside-in" quenching started at approximately 10 kpc from the centre of the galaxy, 625 Myr ago. We therefore find out the $D(4000)$ of a stellar population that is 625 Myr older than the stellar populations of the average PSB. We find this value to be $D(4000)=1.55$. Next, we check what the difference in brightness is between this stellar population and that of a typical disc-dominated star-forming cluster galaxy ($D(4000)=1.2$). We find that a stellar population of solar metallicity with $D(4000)=1.55$ is a factor of 2 fainter than a stellar population of solar metallicity with $D(4000)=1.2$. Therefore, the outskirts of the starting galaxy models will be faded by a factor of 2 in our outside-in disc-fading toy model.

\subsubsection{S\'ersic profiles of the model galaxies}
\label{outsidein_models}

\begin{table}
	\centering
	\caption{Summary of parameters for model galaxies in the outside-in disc-fading toy model. $n$ is S\'ersic index, $b/a$ is axis ratio and P.A. is position angle measured in degrees counter-clockwise. Dashes correspond to parameters that were not fixed to a single value, and left free in the GALFIT fitting process.}
	\label{tab:outsidein_table}
	\begin{tabular}{lccccc}
    	\hline
        Galaxy & $D(4000)$ & Magnitude & $n$ & $b/a$ & P.A.\\
        \hline\hline
        Starting & 1.2 & 22.00 & 1 & 0.62 & 90\\
         \hline
         Faded & 1.4 & - & - & - & -\\
          \hline
	\end{tabular}
\end{table}

\begin{figure*}
	\includegraphics[width=0.6\textwidth]{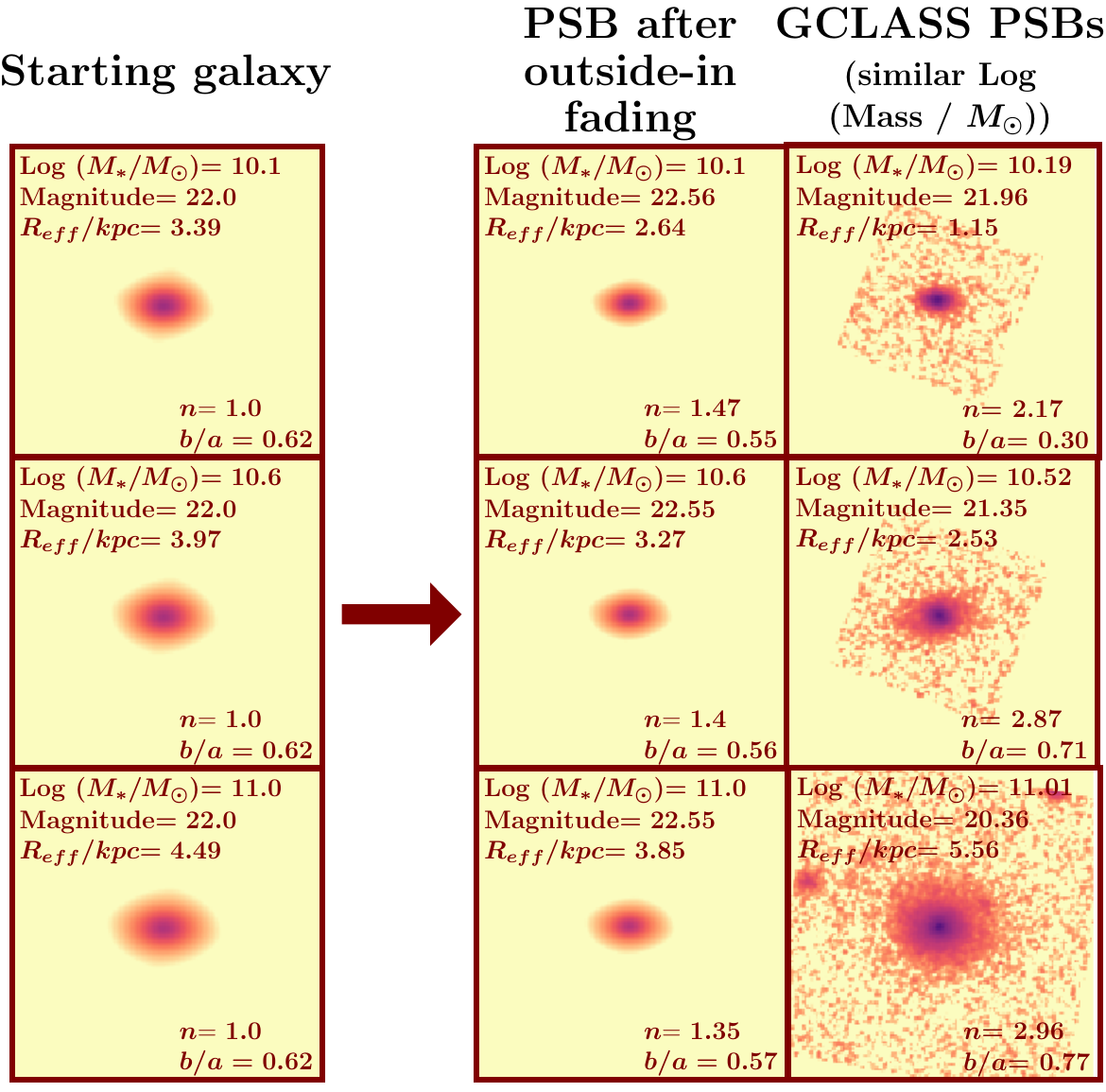}
    \caption{Examples of models created with GALFIT for the outside-in disc-fading toy model. Model galaxies for three stellar masses (Log$(M_{*}/$M$_{\odot})=10.1$, $10.6$ and $11.0$) are shown (first and second column). The starting galaxy models follow the $z~\mathtt{\sim}~1.1$ star-forming mass--size relation in the field. All starting galaxy models have Magnitude$=22.0$, S\'ersic index, $n=1.0$ and axis ratio, $b/a=0.62$. The starting galaxy model is then faded from the outside-in to replicate the stellar populations of the GCLASS PSBs (see text for details). The resulting faded galaxy models are shown in the second column. Final column shows PSBs from GCLASS with similar stellar masses to the galaxy models. Parameters listed in all cutouts show the measured values from GALFIT. For the models, sizes are converted into kpc assuming the models are at $z=1$. The dimensions of all cutouts are the same, as well as the position angle (measured counter-clockwise) of all models and the PSBs, set to 90 degrees. The colourmap is logarithmic.}
    \label{fig:outsidein_fading_models}
\end{figure*}

\begin{figure*}
	\includegraphics[width=\textwidth]{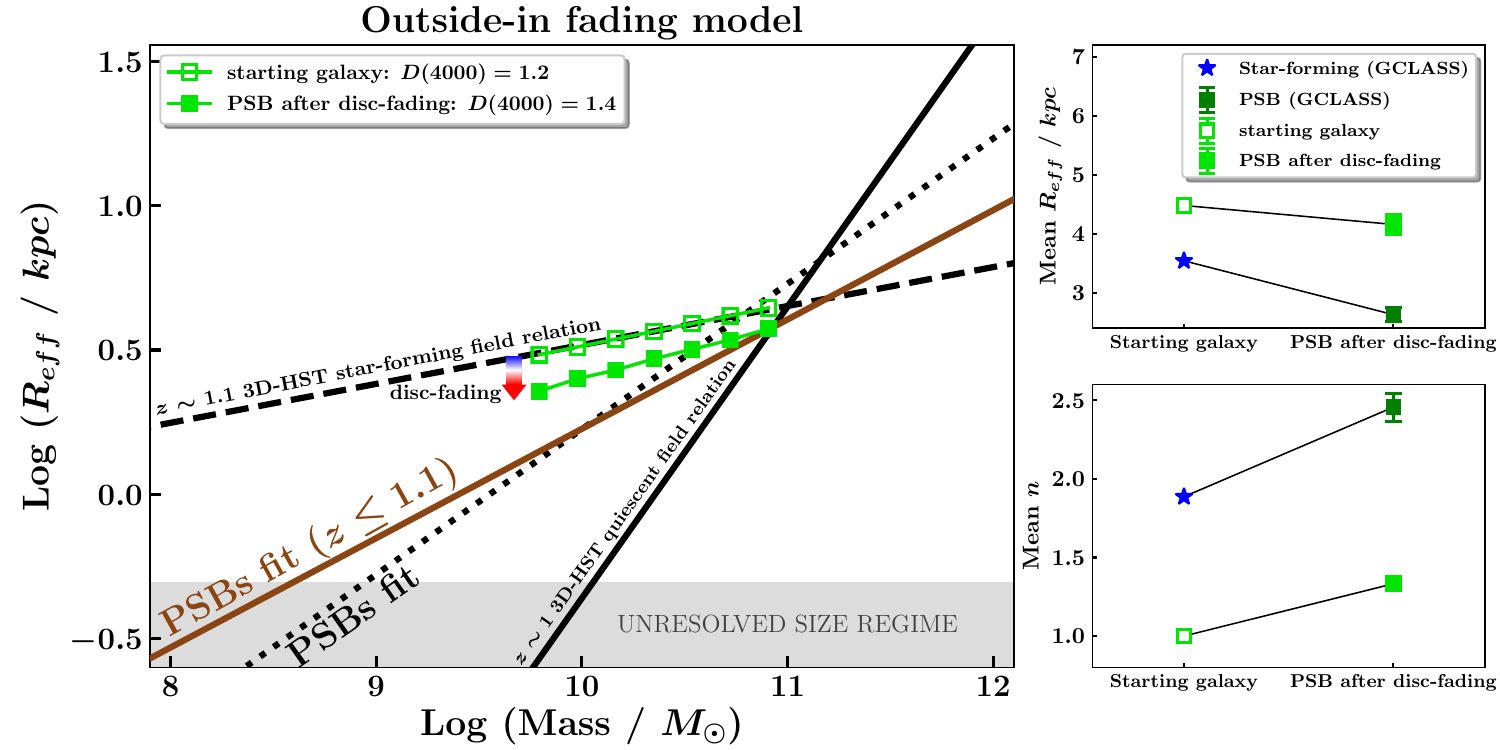}
    \caption{Outside-in disc-fading toy model. Main panel: median size measurements for seven stellar mass bins of the starting (open green squares) and faded (filled green squares) galaxy models. Starting galaxy models follow the $z\sim1.1$ star-forming field mass--size relation (see text for reasoning). Starting galaxy models are faded from the outside-in to replicate the stellar populations of the GCLASS PSBs (see text for details). The region in which size measurements fall below the resolution limit of the GCLASS F140W images is shaded in grey. Top and bottom right panels: average half-light radius and S\'ersic index values for the starting and faded galaxy models along with the values for the GCLASS star-forming and PSB cluster galaxies (see text for more details). Error bars on GCLASS measurements are standard errors in the mean. Error bars on the model results are standard deviations from monte carlo sampling (see text for details).}
    \label{fig:outsidein_fading_toy_model}
\end{figure*}

Starting galaxies are modelled as single component S\'ersic profiles using GALFIT, with structural parameters fixed to those values stated in Table~\ref{tab:outsidein_table}. The F140W magnitude for the starting galaxy models is set to the average magnitude of disc-dominated star-forming cluster galaxies in GCLASS. $n=1$ is the typical S\'ersic index of a disc-dominated star-forming cluster galaxy (e.g. Figure~\ref{fig:morphology}) and the axis ratio is set to the typical axis ratio of a star-forming galaxy at $z\sim1$ \citep{Hill2019}. The PSBs began quenching $\sim0.4$~Gyr ago \citep{Muzzin2014a} and were therefore star-forming galaxies $\sim0.4$~Gyr ago. $\sim0.4$~Gyr prior to $z\sim1$ corresponds to $z\sim1.1$. Therefore, the starting galaxy models have sizes as dictated by the $z\sim1.1$ star-forming field mass--size relation. These models are then convolved with a PSF from the GCLASS F140W images. This is to account for the smearing of images due to the resolution limit of WFC3. This PSF is the same pre-selected star (see \citealt{Matharu2018}) from the WFC3 F140W images of the GCLASS clusters used in the disc and bulge disc-fading toy model. A noise map is not used, since our models are noiseless (apart from the noise introduced by convolving with the PSF) and no sky estimation is carried out since there is no sky background in our models. The same two-GALFIT-run approach described in \cite{Matharu2018} is used to determine the sizes and S\'ersic indices of the starting galaxy models.

We also create starting galaxy models that are not convolved with the PSF. These are used to create the faded galaxy models. The benefit of the unconvolved starting galaxy models is that their pixel values are symmetric about the $x$ and $y$ axes running through the central pixel. This allows us to implement an outside-in fading gradient evenly across the model. As mentioned in Section~\ref{outsidein_fading}, we will be using NGC 4330 as a guide for how outside-in fading operates on a galaxy undergoing ram-pressure stripping. This galaxy has a stellar mass of Log$(M_{*}/$M$_{\odot})=9.8$, and the onset of outside-in quenching begins at $\sim10$~kpc from the centre of the galaxy. We check what size NGC 4330 has as dictated by the low-redshift star-forming mass--size relation\footnote{This relation is calculated in the same way the $z\sim1$ field relations were calculated in \cite{Matharu2018}, except we use all the star-forming galaxies in 3D-HST at $0<z<0.5$, under the fitting criteria. There is also only one sample, so the fitting method is run 1000 times on this sample to capture the range of possible intercept and gradient values. The average value for the intercept and gradient are used for the final relation.}. We find that NGC 4330 would have $R_{eff}=3.2$~kpc in F140W if it followed the low redshift star-forming mass--size relation. Comparing this size to the size at which outside-in quenching begins for this galaxy (10 kpc), we can say that outside-in quenching begins at $\sim3R_{eff}$ for NGC 4330. We use this radial boundary as the boundary at which outside-in fading begins for our model galaxies. Therefore at $3R_{eff}$ from the centre of each unconvolved galaxy model, the pixel values are divided by 2, and the central pixel of each model is divided by 1.5 (see Section~\ref{outsidein_fading} for reasoning). For all pixel values in between, linear interpolation is performed to find the relevant fading factors. The same interpolation grid is extended to beyond $3R_{eff}$ all the way to the edges of each model cutout, such that there is no discontinuity in the fading. These unconvolved faded galaxy models are then convolved with the PSF to create the final faded galaxy models. Again, no noise map is used and no sky background estimation is performed. Examples of galaxy models for three stellar masses within the mass completeness limits of GCLASS (Section~\ref{PSBS_mass_size}) are shown in Figure~\ref{fig:outsidein_fading_models}. Alongside the faded galaxy models, we show select PSBs that have a similar stellar mass to the faded model galaxies.

\subsubsection{The stellar mass--size relation of outside-in faded galaxies}

In the largest panel of Figure~\ref{fig:outsidein_fading_toy_model}, we show the resulting median size measurements for the starting and faded galaxy models for multiple stellar mass bins within the GCLASS star-forming mass completeness limit (Log$(M_{*}/$M$_{\odot})>9.60$). We find that there is a fall in the size of a galaxy at fixed stellar mass due to outside-in fading. Whilst this fall in size is larger than that seen in the disc and bulge disc-fading model (20\% compared to 16\% in the disc and bulge disc-fading model, measured at the median stellar mass of the PSBs at $z\leqslant1.1$ within the quiescent mass completeness limit (Log$(M_{*}/$M$_{\odot}) = 10.39$), it is not enough to explain the stellar mass--size relation of the PSBs.

In the top right-hand panel of Figure~\ref{fig:outsidein_fading_toy_model}, we show the average sizes of starting and faded galaxies which have identical stellar masses to the GCLASS star-forming cluster galaxies beyond the star-forming mass completeness limit and PSBs at $z\leqslant1.1$ with reliable spectroscopic redshifts that are beyond the quiescent mass completeness limit, respectively\footnote{The errors on these measurements come from monte carlo sampling, which is described in Section~\ref{faded_ms}.}. We also show the average sizes of the star-forming galaxies in GCLASS within the star-forming mass completeness limit and PSB galaxies in GCLASS at $z\leqslant1.1$ with reliable spectroscopic redshifts that are beyond the quiescent mass completeness limit, respectively, for comparison. It is evident that just as there is a drop in average size between the star-forming and PSB cluster galaxies, there is also a drop in average size as a result of outside-in fading in the models. The magnitude of this drop is 3 times larger than that observed in the disc and bulge disc-fading model.

\subsubsection{The average S\'ersic index of outside-in faded galaxies}

In the bottom right-hand panel of Figure~\ref{fig:outsidein_fading_toy_model}, we compare the average S\'ersic indices of the starting and faded galaxy models\footnote{For the starting galaxy models, these are calculated from model galaxies that have exactly the same stellar masses as the star-forming cluster galaxies in GCLASS beyond the star-forming mass completeness limit. For faded galaxy models, these are calculated from model galaxies that have exactly the same stellar masses as the PSBs beyond the quiescent mass completeness limit.} to the average S\'ersic indices of the GCLASS star-forming cluster members and the PSBs\footnote{within the star-forming and quiescent mass completeness limits, respectively.}.

We can straight away see that outside-in fading has led to a bulgier morphology on average for the faded galaxy models. The rise in S\'ersic index is approximately 2 times higher in this model compared to in the disc and bulge disc-fading model. However, it is still 42\% smaller than the difference observed in the observations.

\subsection{Disc-fading toy model (bulge and outside-in faded disc)}
\label{bulge_disc_outsidein_model}
\begin{figure*}
	\includegraphics[width=\textwidth]{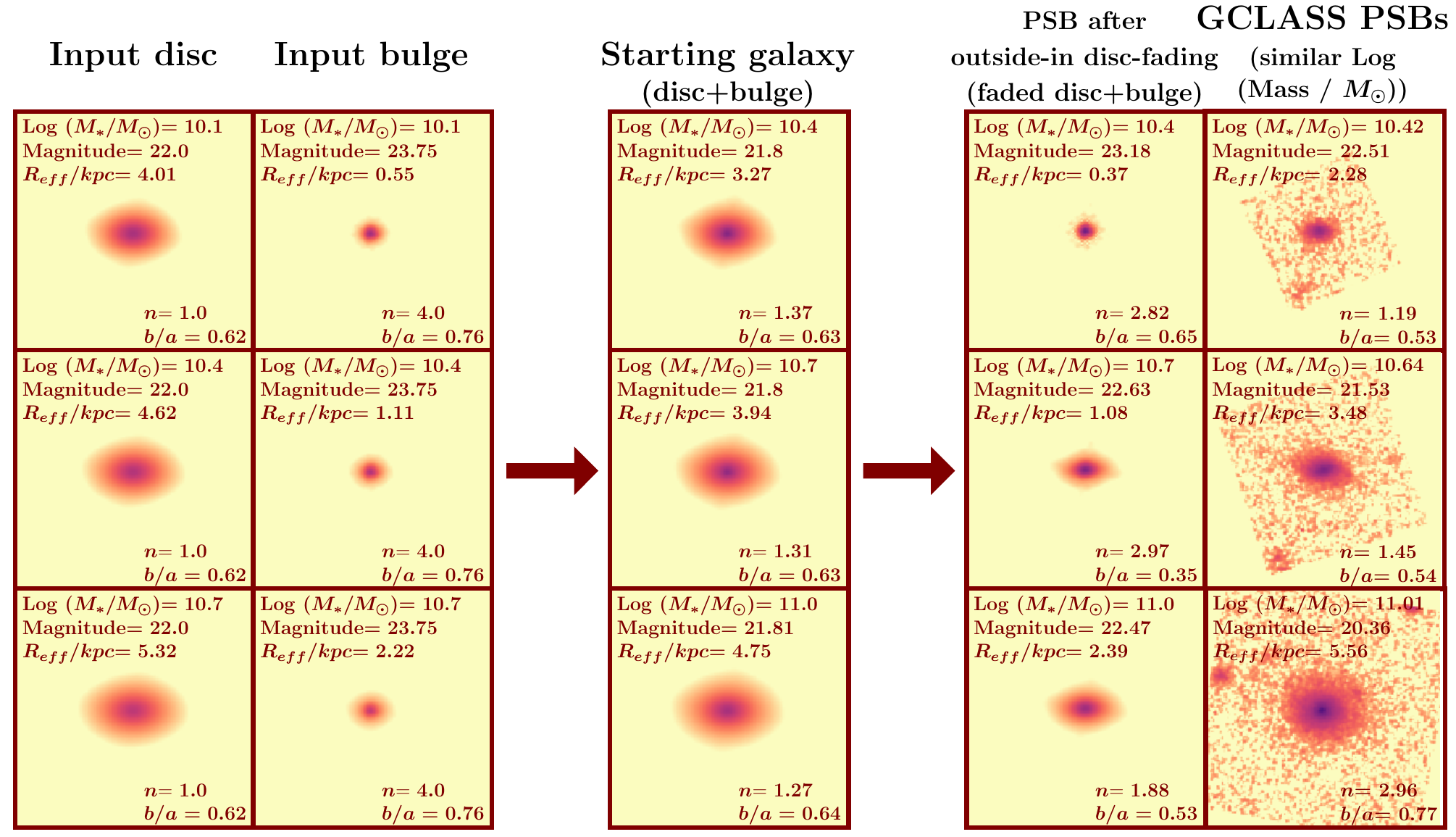}
    \caption{Examples of models created with GALFIT for the bulge + outside-in disc-fading toy model. Model galaxies for three stellar masses (Log$(M_{*}/$M$_{\odot})=10.4, 10.7$ and $11.0$) are shown (third and fourth column). First and second columns show the disc and bulge models, respectively. The discs follow the $z~\mathtt{\sim}~1.1$ star-forming disc mass--size relation in the field. The bulges follow the $z~\mathtt{\sim}~1$ quiescent field mass--size relation (see Section~\ref{profiles}). All disc models have Magnitude$=22.0$, S\'ersic index, $n=1.0$ and axis ratio, $b/a=0.62$. All bulge models have a magnitude that is 5 times fainter than the disc models (see Section~\ref{relative}). All bulge models are set to have $n=4.0$ and $b/a=0.76$. Bulge and disc models have the same stellar masses. The disc and bulge models are then combined to create starting galaxies shown in the third column. The disc model is then faded from the outside-in, as per the details discussed in Section~\ref{bulge_outsidein_disc_fading}, and combined with the original bulge models. The resulting faded galaxies are shown in the penultimate column. Final column shows PSBs from GCLASS with similar stellar masses to the galaxy models. Parameters listed in all cutouts show the measured values from GALFIT. For the models, sizes are converted into kpc assuming the models are at $z=1$. The dimensions of all cutouts are the same, as well as the position angle (measured counter-clockwise) of all models and the PSBs, set to 90 degrees. The colourmap is logarithmic.}
    \label{fig:outsidein_bulge_disc_fading_models}
\end{figure*}

\begin{figure*}
	\includegraphics[width=\textwidth]{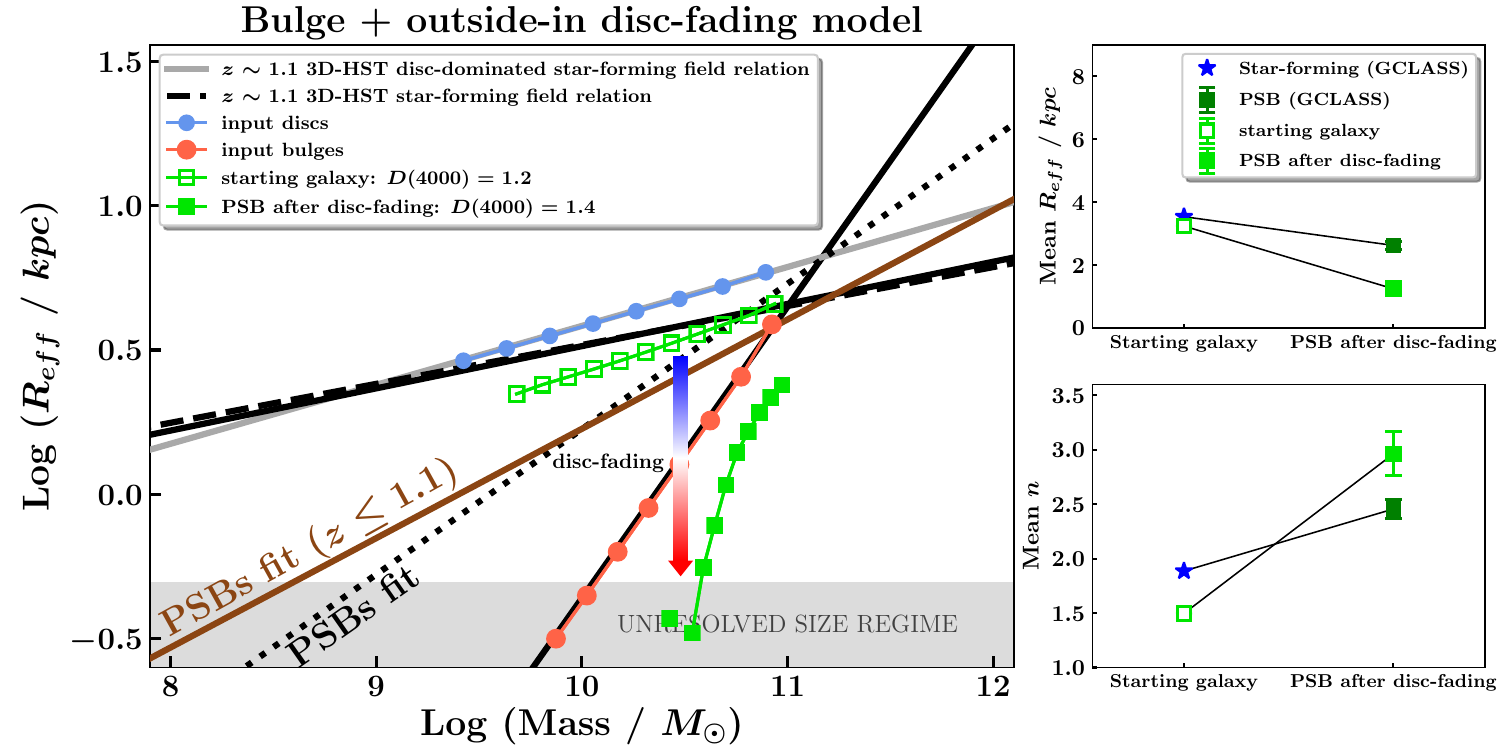}
    \caption{Bulge + outside-in disc-fading toy model. Main panel: median size measurements for 8 stellar mass bins of the disc (small blue circles) and bulge (large red circles) input S\'ersic models that were created to follow the $z\sim1.1$ disc-dominated star-forming field mass--size relation and the $z\sim1$ quiescent field mass--size relation, respectively. For reference, we show the $z\sim1.1$ and $z\sim1$ star-forming field mass--size relations as dashed and solid black lines respectively. Bulges are five times fainter than the discs at fixed stellar mass (see text for reasoning). The starting galaxy is composed of the disc+bulge input models, median size measurements of which in eleven stellar mass bins are shown as open green squares. Corresponding median size measurements of the disc-faded galaxies are shown as filled green squares. Since $M_{*, bulge} = M_{*, disc}$, the stellar masses of the starting and disc-faded galaxies are $2M_{*,bulge}$ (or $2M_{*,disc}$). The disc is faded outside-in such that the stellar populations of the faded galaxy match the stellar populations of the PSBs (see Section~\ref{bulge_outsidein_disc_fading}). The region in which size measurements fall below the resolution limit of the GCLASS F140W images is shaded in grey. Top and bottom right panels: average half-light radius and S\'ersic index values for the starting and disc-faded galaxy models along with the values for the GCLASS star-forming and PSB cluster galaxies (see text for more details). Error bars on GCLASS measurements are standard errors in the mean. Error bars on the model results are standard deviations from monte carlo sampling (see text for details).}
    \label{fig:bulge_disc_outsidein_fading_toy_model}
\end{figure*}

In Section~\ref{outsidein_model}, we presented a more representative model of disc-fading, based on recent observations of how ram-pressure stripping operates in low redshift cluster galaxies. However, we did not model our galaxies with a bulge and disc component, as was done in the uniform disc-fading model presented in Section~\ref{model2}. The natural question therefore arises, whether modelling a galaxy with a bulge and disc component, {\it and} fading the disc with the more representative outside-in fading, can produce galaxies with structural and stellar population properties similar to the PSBs.

In this Section, we therefore present a disc-fading model that attempts to combine representative modelling of a galaxy's structure and the fading process. The bulge and disc models are constructed in exactly the same way as those in the bulge and disc disc-fading model (Section~\ref{model2}). The difference between the model presented in this Section and the previous two models is how the fading is implemented. The allowed amount of fading in this model remains constrained by the $D(4000)$ measurements of the star-forming, PSB and quiescent cluster galaxies in GCLASS.

\subsubsection{Outside-in disc fading in the bulge+disc case}
\label{bulge_outsidein_disc_fading}

As was highlighted in Section~\ref{relative}, there are strict boundaries set on how much the disc can be faded in a bulge+disc model. These boundaries are dictated by the spectroscopic properties of the cluster galaxies. In the case of uniform disc-fading, the disc was faded by a factor of 2 such that the resulting composite bulge+disc model would represent a galaxy that had similar stellar populations to the PSBs. In the model presented in this Section, we want to fade the disc ``outside-in", yet still produce a galaxy with similar stellar populations to the PSBs. To do this, we need to ensure that the outside-in fading we implement on the disc components results in the same total brightness as when we fade the same discs uniformly by a factor of 2.

The outside-in fading is implemented using the same method described in Section~\ref{outsidein_models} with a few modifications. For each model disc, we first check what the total brightness of the disc is when it is uniformly faded by a factor of 2. We do not fade the central pixel of the disc model. We check how much the disc model needs to be faded at $\sim3R_{eff}$ (see Section~\ref{outsidein_fading} for reasoning) from the centre, such that the entire model has the same brightness as when it is uniformly faded by a factor of 2. We then adjust the slope of our linearly interpolated fading accordingly. Examples of galaxy models for three stellar masses within the mass completeness limits of GCLASS are shown in Figure~\ref{fig:outsidein_bulge_disc_fading_models}. Alongside the faded galaxy models, we show select PSBs that have similar stellar masses to the model galaxies.

\subsubsection{The stellar mass--size relation of bulges + outside-in faded discs}
\label{mass_size_bulge_outsidein_disc}

In the largest panel of Figure~\ref{fig:bulge_disc_outsidein_fading_toy_model}, we show the median size measurements for the starting (open green squares) and faded (filled green squares) galaxy models for multiple stellar mass bins, as well as the disc (small blue circles) and bulge (large red circles) components. The lines are the same as in Figure~\ref{fig:fading_toy_model}. The most striking aspect of these measurements is that this model of disc-fading leads to a dramatic drop in the size of galaxies. So much so, that the half-light radii are smaller than those of the bulge models. At the median stellar mass of the PSBs with reliable spectroscopic redshifts at $z\leqslant1.1$ within the quiescent mass completeness limit (Log$(M_{*}/$M$_{\odot}) = 10.39$), the half-light radius shrinks by 89\%. This large size reduction is likely due to the fact that the central pixels of the disc models have not been faded. As a result, the outside-in fading implemented has a particularly steep gradient. Consequently, there is far more light in the central regions of the faded galaxies compared to the corresponding bulge model. This is also evident in the first row of Figure~\ref{fig:outsidein_bulge_disc_fading_models}. This leads to a galaxy with a particularly steep light profile, and therefore a smaller half-light radius.

In the top right-hand panel of Figure~\ref{fig:bulge_disc_outsidein_fading_toy_model}, we show the average sizes of the starting and faded galaxy models for models that have stellar masses identical to the GCLASS star-forming cluster galaxies and $z\leqslant1.1$ PSBs with reliable spectroscopic redshifts beyond the star-forming and quiescent mass completeness limits, respectively. We also show the average sizes of the star-forming and $z\leqslant1.1$ PSB (with reliable spectroscopic redshifts) cluster galaxies in GCLASS within the star-forming and quiescent mass completeness limits, respectively. Errors are calculated in the same way as for the previous disc-fading models\footnote{If error bars are not visible, it is because the error bar is smaller than the size of the marker.}. The dramatic reduction in the average size of a galaxy is evident in this model, with the faded models reaching an average size that is smaller than half that of the average PSB.

\subsubsection{The average S\'ersic index of bulges + outside-in faded discs}

In the bottom right-hand panel of Figure~\ref{fig:bulge_disc_outsidein_fading_toy_model}, we compare the average S\'ersic indices of the starting and faded galaxy models to the average S\'ersic indices of the star-forming and PSB cluster galaxies\footnote{These, as well as their errors are calculated in the same way as in the first two disc-fading toy models.}. It is immediately noticeable that this disc-fading model -- unlike the previous two models -- can produce more than the required rise in S\'ersic index that is seen in the observations. This means that with more sophisticated modelling, the bulge plus outside-in faded disc model has the ability to reproduce the required level of increase in S\'ersic index that is seen in the observations.

\subsection{Caveats}

It should be emphasised that while considerable effort has been made to model disc-fading realistically, the three toy models presented in Sections~\ref{model2}, \ref{outsidein_model} and \ref{bulge_disc_outsidein_model} are still very simplistic. For the models that consider bulge and disc components, only the case where the two components have the same stellar mass has been considered. In all three of the models presented, we only consider models with specific S\'ersic indices, axis ratios, F140W magnitudes and position angles. Realistically, galaxies will have a range of disc-to-bulge stellar mass ratios, S\'ersic indices, axis ratios, F140W magnitudes and position angles. The fading of galactic discs would also occur in varying amounts and the stellar populations would vary. These variances would cause a scattering in the positions of the starting and faded galaxies on the mass--size plane.

Furthermore, our understanding of outside-in fading, and its implementation in Sections \ref{outsidein_model} and \ref{bulge_disc_outsidein_model}, is based on a single -- albeit well-studied -- example of a cluster galaxy undergoing ram-pressure stripping at low redshift (see Section~\ref{outsidein_fading}). Ram-pressure stripping may operate very differently at high redshift, where gas fractions are higher than at low redshift. There may also be a stellar mass dependency on how ram-pressure stripping operates, which we will discuss in more detail in Section~\ref{discussion}.

Particularly in the last disc-fading model (Section~\ref{bulge_disc_outsidein_model}), the outside-in fading seems to be too aggressive. The way in which the fading is implemented leads to a large contraction in the half-light radius of a galaxy. Certainly, if the gradient of the outside-in fading was shallower, the magnitude of this contraction could be reduced (as is evident in Figure~\ref{fig:outsidein_fading_toy_model} for example), and faded galaxies that approximately follow the PSB mass--size relation could be achieved.

Nevertheless, it is clear that with more sophisticated modelling -- particularly in the exploration of different fading gradients -- outside-in fading can lead to the required contraction in size {\it and} the required increase in bulge-dominance that is observed between the star-forming and PSB cluster galaxies.

\section{DISCUSSION}
\label{discussion}

\subsection{Evidence for the PSBs being transition galaxies}

The PSBs are cluster galaxies that were selected as being recently quenched -- and therefore intermediary to star-forming and quiescent galaxies -- based solely on their spectroscopic properties in the optical (Section~\ref{PSB_sample}). In Sections~\ref{PSBS_mass_size} and \ref{morphology_sec}, we presented two pieces of evidence that suggest the PSBs also have structural properties intermediary to those exhibited by typical star-forming and quiescent cluster galaxies: 1) The PSBs lie on a distinct stellar mass--size relation that almost perfectly bisects the region between the star-forming and quiescent mass--size relations, and 2) Morphologically, out of the three populations (Star-forming, PSB and Quiescent), the PSB population contains the largest fraction of intermediate-type galaxies. The fraction of bulge-like galaxies in the PSB population is also intermediary to the fraction of bulge-like galaxies found in the star-forming and quiescent populations. The fact that the PSBs all lie on the same mass--size relation suggests that most, if not all of them formed via the same physical process. After inspecting the HST WFC3 F140W imaging of the PSBs (Section~\ref{direct_images}), we find most (20 out of 23) of them have symmetrical, undisturbed morphologies in the stellar continuum. This observation rules out physical quenching mechanisms such as mergers, harassment \citep{Moore1996,Moore1998} or tidal stripping which lead to the redistribution of stars. Instead, it supports a scenario where the PSBs are formed via a quenching process that only effects the gas reservoir of the galaxy.

Quenching mechanisms prevalent in the cluster environment which solely affect the gas content of a galaxy include ram-pressure stripping \citep{Gunn1972} and strangulation \citep{Larson1980}. Spectacular examples of on-going ram-pressure stripping quenching galaxies rapidly have been observed in low redshift clusters (e.g. \citealt{Abramson2011a,Ebeling2014,Bellhouse2017,Gullieuszik2017,Sheen2017,Fossati2018,Cramer2019}). Both the rapidity of this process (see also \citealt{Boselli2016a}) and its potential to indirectly reduce the half-light radius of a galaxy support the short quenching timescales found for the PSBs from our previous work \citep{Muzzin2014a} and the location of the PSB mass--size relation relative to the star-forming one in this study.

It must be noted that there is no definitive way to identify galaxies that are transitioning from star-forming to quiescent. Therefore, results on galaxies that are similarly selected to the GCLASS PSBs tend to vary, making comparison with the literature challenging. In general, PSBs have strong Balmer absorption and no emission lines which are induced by star formation. However, by altering the exact spectroscopic selection criteria, very different samples of transition galaxies can be selected which exhibit a range of physical properties (e.g. \citealt{Lemaux2017,Wilkinson2017}). Some of these selection criteria can lead to significant contamination by dusty star-forming galaxies \citep{Wu2014}. Particularly strong Balmer absorption can be indicative of a starburst occurring in the galaxy prior to quenching (Section~\ref{PSB_sample}). Such starbursts are likely to be the result of violent interactions such as galaxy-galaxy mergers or harassment, not gas-related processes such as ram-pressure stripping or starvation (e.g. \citealt{Wild2009}).

Nevertheless, many other similar works to ours have also concluded that PSB galaxies in clusters are most likely the product of ram-pressure stripping \citep{Poggianti2009, Pino2014, Fritz2014, Paccagnella2019}. In most cases, this is usually associated with the higher occurence of PSBs in clusters compared to other environments such as the field and groups \citep{Ma,Poggianti2009,Vergani2009,Muzzin2012,Wu2014,Galametz2018,Paccagnella2019}, suggesting cluster-specific quenching mechanisms are more efficient in transforming star-forming galaxies into PSBs. More detailed studies of the spatial distribution of PSBs in clusters have uncovered more convincing evidence for ram-pressure stripping being responsible for their emergence. \cite{Ma} found that their E+A cluster galaxies were evenly distributed within the projected ram-pressure stripping radius of the cluster MACS J0717.5+3745. Going a step further, \cite{Fritz2014} found that their PSBs were radially distributed in a shell-like structure, congregating in between where most of the star-forming and quiescent cluster galaxies were found. This seemed to suggest a transition region in clusters, where there is a change in the physical properties, perhaps in the density of the intracluster medium, triggering efficient ram-pressure stripping. Similarly, \cite{Muzzin2014a} found that most of the GCLASS PSBs with spectroscopic redshifts from GMOS followed a coherent ``ring" in clustercentric velocity versus position phase space. Most of the GCLASS PSBs  are situated at small clustercentric radii ($< 0.5 R_{200}$) with high clustercentric velocities. Such conditions were found to be consistent with the radii at which ram-pressure stripping and starvation are most effective in clusters \citep{Treu2003, Moran2007,Bahe2013}.

Since ram-pressure stripping is capable of removing the cold gas reservoir of a galaxy, it can remove the fuel for star formation. With no younger stars actively forming, the average age of the stellar populations in the galaxy increases, eventually leading to a fall in the brightness of the galaxy. Ram-pressure stripping is more effective at removing gas from galactic discs \citep{Abadi1999}. With a strong body of evidence suggesting the GCLASS PSBs are a product of ram-pressure stripping, we decided to test whether the fading of a star-forming galaxy's disc could lead to a population of galaxies that would lie along the PSB mass--size relation. 

\subsection{Outside-in disc-fading qualitatively explains the structural properties of the PSBs}

Initially, we modelled star-forming galaxies as having both a disc and bulge component of equal stellar mass. The disc would then uniformly fade such that the galaxy's stellar populations matched those of the PSBs (Section~\ref{model2}). In this scenario, we found that the fading did not produce model galaxies with similar sizes and S\'ersic indices to the average PSB (Figure~\ref{fig:fading_toy_model}). Far more fading was required to explain the small sizes and high S\'ersic indices of the PSBs. This is perhaps unsurprising, since observational evidence suggests quenching in clusters does not operate in this way. 

Given that numerous previous results suggest that outside-in quenching occurs in low redshift clusters, we therefore proceeded to model outside-in fading on the mass--size plane (Section~\ref{outsidein_model}), finding that it also fails to produce model galaxies with similar structural properties to the PSBs. Furthermore, this model fails to reproduce the steep slope of the PSB stellar mass--size relation. A possible reason for this could be due to a difference in the fading rate for low- and high-mass galaxies. Outside-in fading may operate at a faster rate for low-mass galaxies, but much slower for high-mass galaxies. This may also allow for the average S\'ersic index to rise, since at fixed stellar mass, all the galaxies with Log$(M_{*}/M_{\odot})\lesssim10.9$ --which are in the majority -- will have smaller sizes at fixed stellar mass, more similar to quiescent galaxies.

Nevertheless, we decided to combine our first and second attempts at modelling disc-fading into a third model (Section~\ref{bulge_disc_outsidein_model}). From the first model (Section~\ref{model2}), we took the more realistic modelling of galactic structure, modelling our galaxies with a disc and bulge component. From the second model (Section~\ref{outsidein_model}), we took the more realistic modelling of ``outside-in" fading. The result is a model that can explain the high S\'ersic indices of the PSBs and has the potential to produce disc-faded galaxies that approximately follow the PSB mass--size relation. The results of these three disc-fading toy models have shown that the PSBs are plausibly the product of star-forming galaxies which have experienced outside-in disc-fading. This outside-in disc-fading is likely to have been brought on by quenching processes such as ram-pressure stripping. To better reproduce the structural properties of the PSBs with outside-in disc-fading, various fading gradients with more sophisticated modelling need to be explored. The gradient of outside-in fading could very much rely upon the stellar mass of the galaxy, which in turn dictates the slope of the relation the faded galaxies lie on. This therefore means the outside-in fading models presented in Sections \ref{outsidein_model} and \ref{bulge_disc_outsidein_model} approximately represent two extremes in fading gradients.

Many recent works have found that disc fading could explain the formation of transition galaxies in clusters. Using a bulge and disc decomposition of quiescent galaxies in the Coma cluster, \cite{Head2014} find that S0 (intermediate-type) galaxies have systematically bluer bulges and discs further away from the cluster centre. This observation favoured environmentally-induced disc fading as the dominant route for S0 formation. Going further, \cite{Bedregal2011} were able to show that most S0 galaxies in the Fornax cluster have quenched ``outside-in", with the latest star formation in the central regions being capable of enhancing the brightness of the bulge. Similarly, \cite{Jaffe2011} find that kinematically-disturbed emission-line cluster galaxies out to $z\sim1$ have truncated gas discs and that their star formation is more centrally concentrated than analogous galaxies in low density environments. They suggest that gas is either more efficiently removed from the outskirts of galaxies and/or is driven towards the centre. Driving gas towards the centre of the galaxy would further fuel the centrally-concentrated star formation and contribute to the build-up of a significant bulge. Perhaps this may be an attribute of gas-stripping processes which was not considered in our toy models. If ram-pressure stripping could also contribute to enhancing the stellar mass build-up of the bulge, a scenario could emerge where there is bulge {\it brightening} as well as disc fading. This could most certainly explain the more bulgier morphologies of the PSBs compared to the faded galaxies in our fading models. Enhanced star formation in the core of a ram-pressure stripped galaxy in the Coma cluster was seen in \cite{Cramer2019} for example. Furthermore, \cite{Fraser-McKelvie2018} found that the bulges of low mass ($M_{*}<10^{10}$M$_{\odot}$) S0 galaxies in the MaNGA survey hosted younger stellar populations than their discs, supporting an outside-in fading scenario where the bulge may end up brighter than the disc.

Nonetheless, other works have stated that outside-in quenching may not be the only explanation for how cluster galaxies quench. Using a bulge-disc decomposition of late- and early-type galaxies in six low redshift clusters, \cite{Christlein2004} found that the bright end of the disc luminosity function does not vary much with bulge fraction, but the bright end of the bulge luminosity function increases with bulge fraction. Quenching in clusters seems to favour the bulge becoming brighter rather than the disc fading as a galaxy moves from spiral to elliptical morphology. This seems to suggest that environmental quenching processes are not enough to explain the structural properties of recently quenched cluster galaxies. For example, to explain how low mass (Log$(M_{*}/M_{\odot})<10.5$) PSBs evolved from compact star-forming galaxies in $z<1$ clusters, \cite{Socolovsky2018} found that a combination of gas stripping and strong outflows from stellar or AGN feedback were needed to explain their structural properties.

\subsection{Recently quenched galaxies \& evolution in the quiescent mass--size relation with redshift}

The distinct location of the PSBs on the mass--size plane has important implications for the growth in the average size of quiescent cluster galaxies with decreasing redshift. The majority of the PSBs lie on the large size end of the quiescent distribution at fixed stellar mass (see Figure~\ref{fig:mass_size_PSBs}). Their location in Figure~\ref{fig:mass_size_PSBs} provides direct evidence showing that galaxies which quench later are on average larger in size than galaxies that quenched earlier. Therefore, the addition of recently quenched galaxies to the quiescent population can induce an increase in the average size of quiescent galaxies with decreasing redshift.

In \cite{Matharu2018}, we found that minor mergers could explain the size growth observed in quiescent field galaxies with decreasing redshift at fixed stellar mass. Now, with the study presented in this paper, we confirm that the addition of recently quenched galaxies to the quiescent population will also induce a rise in the average size of quiescent galaxies with decreasing redshift. The relative importance of these two size growth mechanisms however still needs to be determined. For example, a higher occurence of PSBs in clusters have been found by a number of other studies \citep{Poggianti2009,Vergani2009,Muzzin2012,Wu2014, Paccagnella2019}. Since minor mergers are rare between satellite cluster galaxies, these findings may suggest that the addition of recently quenched galaxies to the quiescent population is more important for the size growth observed in the quiescent {\it cluster} population than for the size growth observed in the quiescent {\it field} population with decreasing redshift.

On the other hand, \cite{Carollo2013} claim that the addition of recently quenched galaxies to the quiescent population can explain the majority of the size growth observed in the median size of quiescent field galaxies between $0.2<z<1$. This conclusion was motivated by the fact that these authors do not observe a decrease in the number density of compact quiescent galaxies in this redshift range (which, if observed, would have suggested they have grown physically in size). Since this result, many other studies have also found that progenitor bias plays an important role in the size growth of quiescent galaxies at fixed stellar mass \citep{Cassata2013,Krogager2014,Damjanov2015,Zahid2016,Charbonnier2017,Tacchella2017,Abramson2018,Damjanov2018}. In further support of this conclusion, \cite{Fagioli2016} found that large quiescent field galaxies with $10.5<$~Log$(M_{*}/M_{\odot})<11$ at $0.2<z<0.8$ were composed of younger stellar populations than smaller quiescent field galaxies within the same stellar mass range. Similar results to this one have also been found by \cite{Delaye2014}, \cite{Gargiulo2017}, \cite{Scott2017}, \cite{Williams2017}, \cite{Zahid2017} and \cite{Wu2018}. \cite{VanderWel2014} however did observe a decrease in the number density of compact quiescent galaxies with a much larger sample of field galaxies between $0<z<2$, concluding that minor mergers drive the observed growth in the average size of quiescent galaxies with decreasing redshift. Many other works have also concluded that minor mergers are the main drivers of size growth at fixed stellar mass for the quiescent population \citep{Morishita2014,Damjanov2014,VanDokkum2015,Chan2016,Zanella2016,Oldham2017,Kubo2018}. Regardless, there remains disagreement in which of the two processes dominates the observed size growth \citep{Belli2013,Rutkowski2014,Keating2014,Belli2015}.

Using the results in \cite{Matharu2018} and this work, we confirm that both minor mergers and recently quenched galaxies are responsible for the observed growth in the average size of quiescent galaxies at fixed stellar mass with decreasing redshift. However, better sample statistics spread over a large range in redshift and environments will be required to quantify which of the two processes dominates.

\section{Summary}
\label{summary}
In this paper, we have studied the structural properties of 23 spectroscopically identified recently quenched (or ``poststarburst" (PSB)) cluster galaxies residing in nine clusters at $z\sim1$, to better understand the physical process responsible for their quenching. Our main conclusions are as follows:

\begin{enumerate}
  \item In most cases, the F140W direct images of the recently quenched galaxies show no disturbances to their stellar light profiles that could be the result of violent interactions such as mergers, harassment or tidal stripping. This suggests the quenching mechanism responsible effects solely the gas content of these galaxies.
  \item By fitting the F140W direct images of the recently quenched galaxies with single component S\'ersic profiles, we find that most of the recently quenched galaxies lie on a distinct mass--size relation that almost perfectly bisects the region lying in between the star-forming and quiescent field stellar mass--size relations. This suggests the quenching mechanism responsible is capable of altering the light profile of these galaxies in a very specific way, with the likelihood that most of the recently quenched galaxies underwent the same quenching process.
  \item Using S\'ersic index as a proxy for morphology, we compare the S\'ersic profiles of the recently quenched, star-forming and quiescent cluster galaxies. The highest fraction of intermediate-type galaxies is found in the recently quenched population. Recently quenched galaxies also exhibit a higher fraction of bulge-dominated galaxies than the star-forming population, but a lower fraction compared to the quiescent population. Therefore, recently quenched galaxies have morphologies that are intermediary to those of the star-forming and quiescent populations. However, on average, their morphologies are more similar to those exhibited by the quiescent population.
  \item By modelling a star-forming cluster galaxy with a bulge and disc component of equal stellar mass, we explored how star-forming galaxies move across the mass--size plane as their discs fade uniformly, resulting in galaxies with similar stellar populations to the recently quenched galaxies. Although a reduction in half-light radius and a rise in S\'ersic index is found, the magnitude of these changes in physical properties are too small to explain the structural properties of the PSBs.
  \item We then used a more realistic model of outside-in fading based on recent observations of quenching in clusters. We found that this model still failed to produce galaxies with similar structural properties to the PSBs. A modest drop in average size and rise in average S\'ersic index is achieved, but the steep slope of the PSB mass--size relation was not achieved. This suggests longer fading timescales with increasing stellar mass are required and/or bulge brightening also occurs as part of quenching processes like ram-pressure stripping which could drive gas towards the centre.
  \item By combining the more realistic modelling of a galaxy with a disc and bulge component, and the more realistic outside-in disc-fading, we showed that outside-in disc-fading can explain the high S\'ersic indices of the PSBs and has the potential to produce galaxies that follow the PSB mass--size relation. For both outside-in fading models, further exploration of outside-in fading gradients combined with more sophisticated modelling is required to fully reproduce the structural properties of the PSBs.
  \item The fact that the PSB mass--size relation runs through the large size end of the quiescent distribution of galaxies provides direct evidence for recently quenched galaxies contributing to the size growth of quiescent galaxies with decreasing redshift.
\end{enumerate}

In tandem with the results presented in \cite{Matharu2018}, we confirm both minor mergers and the addition of recently quenched galaxies to the quiescent population are responsible for the rapid size growth at fixed stellar mass observed in the quiescent population of galaxies with decreasing redshift. However, it remains unclear which of these two processes dominate. Determining this will require better sample statistics over a broad redshift range in a variety of environments.

\section*{Acknowledgements}

J.M. is supported by the Science and Technology Facilities Council (STFC). This work is supported by the National Science Foundation through grant AST-1517863, by HST program number GO-15294, and by grant number 80NSSC17K0019 issued through the NASA Astrophysics Data Analysis Program (ADAP). Support for program number GO-15294 was provided by NASA through a grant from the Space Telescope Science Institute, which is operated by the Association of Universities for Research in Astronomy, Incorporated, under NASA contract NAS5-26555. G.B. acknowledges funding of the Cosmic Dawn Center by the Danish National  Research  Foundation. R.D. gratefully acknowledges the support provided by the BASAL Center for Astrophysics
and Associated Technologies (CATA) grant AFB-170002.


\bibliographystyle{mnras}
\bibliography{library}



\appendix

\section{GALFIT results for the PSBs}
\label{GALFIT_psbs}

\begin{figure*}
	\includegraphics[width=\textwidth]{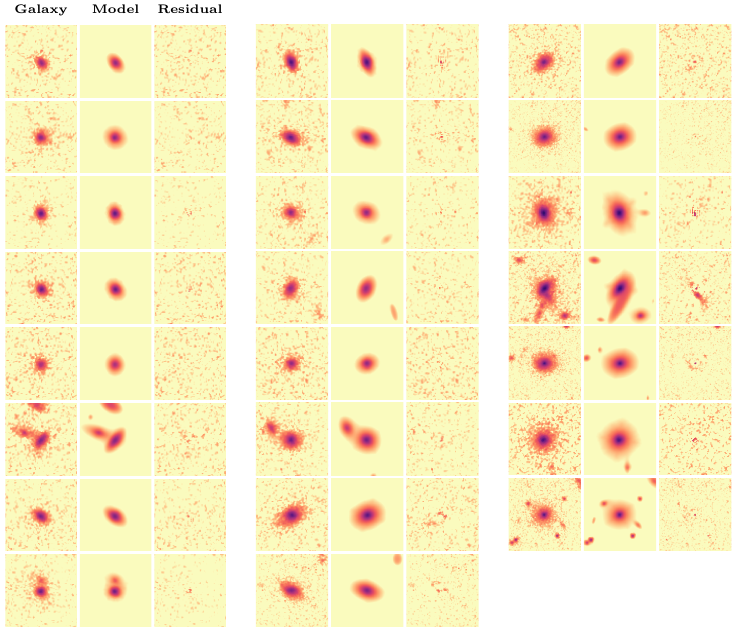}
    \caption{GALFIT fits for all 23 PSBs, ordered by ascending stellar mass (reading from top-to-bottom in each column, starting from the leftmost column). Each of the three columns shows the galaxy cutout, followed by the single-component S\'ersic model and the residual of the fit. The colourmap is logarithmic.}
    \label{fig:GALFIT_PSBs_fig}
\end{figure*}

Figure~\ref{fig:GALFIT_PSBs_fig} shows the GALFIT fits for all 23 PSBs studied in this paper. Higher order structure that can be suggestive of harassment, merger or tidal stripping is absent from most of the residuals. This observation supports our hypothesis made in Section~\ref{direct_images} that the majority of the PSBs likely formed via gas removal processes such as ram-pressure stripping or strangulation.

\section{F140W mass-to-light ratios as a function of D(4000)}
\label{ml_ratios}

\begin{figure}
	\includegraphics[width=\columnwidth]{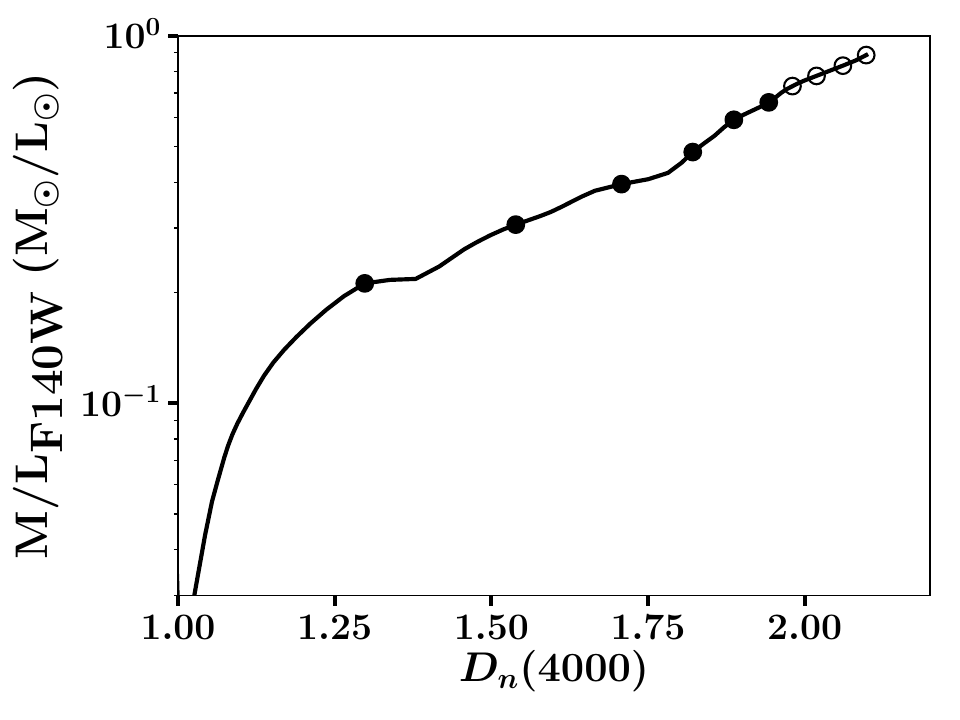}
    \caption{F140W stellar mass-to-light ratios as they would be observed at $z\sim1$ as a function of $D(4000)$ for a set of stellar population models with solar metallicity, an exponentially declining star formation history with $\tau=0.3$~Gyr and a \protect\cite{Chabrier2003} initial mass function using the \protect\cite{Bruzual2003} libraries. Points mark the mass-to-light ratios at intervals of $1$ Gyr. Ages $> 6$~Gyr (open points) are forbidden by the age of the Universe at $z\sim1$.}
    \label{fig:ml_ratios_plot}
\end{figure}

Figure~\ref{fig:ml_ratios_plot} shows how the F140W stellar mass-to-light ratio varies with $D(4000)$ for a set of stellar population models with solar metallicity and a \cite{Chabrier2003} initial mass function using the \cite{Bruzual2003} libraries. An exponentially declining star formation history (SFR~$\propto~\mathrm{e}^{-t/\tau}$) with a star formation timescale, $\tau=0.3$~Gyr is used. We use this particular SFR model because this model was found to be the best fit model to the distribution of $D(4000)$ in GCLASS (see Figure 3 in \citealt{Muzzin2012}).

\bsp	
\label{lastpage}
\end{document}